\documentclass[12pt]{article}

\usepackage{PRIMEarxiv}
\usepackage{algorithm}
\usepackage{algpseudocode}
\usepackage[utf8]{inputenc} 
\usepackage[T1]{fontenc}    
\usepackage{hyperref}       
\usepackage{url}            
\usepackage{booktabs}       
\usepackage{amsfonts}       
\usepackage{nicefrac}       
\usepackage{microtype}      
\usepackage{lipsum}
\usepackage{fancyhdr}       
\usepackage{graphicx}       
\graphicspath{{media/}}     
\usepackage{amsmath}
\usepackage{float}
\usepackage[utf8]{inputenc}
\usepackage{subcaption} 
\usepackage{multirow}%
\usepackage{multicol}

\title{A Physics-Aware Variational Graph Autoencoder for Joint Modal Identification with Uncertainty Quantification}

\author{
  Bhargav P. Nath \\
  Department of Civil Engineering,\\
  Indian Institute of Technology (BHU), \\
  Varanasi, India\\
  \texttt{bhargav.pratimnath.civ21@itbhu.ac.in} \\
   \And
  Mehulkumar Lakhadive \\
  Department of Civil Engineering,\\
  Indian Institute of Technology (BHU), \\
  Varanasi, India\\
  \texttt{lmehulkumar.rajkumar.civ21@itbhu.ac.in} \\
   \And
  Anshu Sharma \\
  Department of Civil \& Environmental Engineering,\\
   University of Strathclyde, \\
   Glasgow, UK\\
  \texttt{anshu.sharma@strath.ac.uk} \\
   \And  
  Basuraj Bhowmik* \\
  Department of Civil \& Environmental Engineering,\\
  University of Strathclyde, \\
  Glasgow, UK\\
  \texttt{basuraj.bhowmik@strath.ac.uk} \\
}

\begin{document}
\maketitle

\begin{abstract}
Reliable modal identification from output-only vibration data remains a
challenging problem under measurement noise, sparse sensing, and
structural variability. These challenges intensify when global modal
quantities and spatially distributed mode shapes must be estimated
jointly from frequency-domain data. This work presents a
physics-aware variational graph autoencoder, termed UResVGAE, for
joint modal identification with uncertainty quantification from power
spectral density (PSD) representations of truss structures. The
framework represents each structure as a graph in which node
attributes encode PSD and geometric information, while edges capture
structural connectivity. A residual GraphSAGE-based encoder,
attention-driven graph pooling, and a variational latent
representation are combined to learn both graph-level and node-level
modal information within a single, unified formulation. Natural
frequencies and damping ratios are predicted through evidential
regression, and full-field mode shapes are reconstructed through a
dedicated node-level decoder that fuses global latent information
with local graph features. Physical consistency is promoted via
mode-shape reconstruction and orthogonality regularisation. The
framework is assessed on numerically generated truss populations
under varying signal-to-noise ratios and sensor availability.
Results demonstrate accurate prediction of natural frequencies,
damping ratios, and mode shapes, with high modal assurance criterion
values and stable performance under noisy and sparse sensing
conditions. Reliability analysis indicates that the predictive
uncertainty is broadly consistent with empirical coverage. The
proposed framework offers a coherent and physically grounded
graph-based route for joint modal identification with calibrated
uncertainty from frequency-domain structural response data.
\end{abstract}


\section{INTRODUCTION}

Accurate identification of modal parameters from measured structural
responses is a central objective in structural health monitoring (SHM),
yet it remains fundamentally difficult in practice owing to unknown
excitation, measurement noise, and spatially sparse sensing
configurations \cite{reynders2012system}. Under
operational conditions, these factors introduce substantial uncertainty
into the estimation of natural frequencies, damping ratios, and mode
shapes \cite{doebling1996damage}. Such difficulties are especially
acute when frequency-domain representations such as power spectral
density (PSD) are employed, since high-dimensional inputs and noise
contamination can hinder reliable modal extraction. Higher-order and
closely spaced modes are particularly vulnerable to these effects, as
noise contamination and limited spectral resolution in PSD
representations can reduce identification accuracy
\cite{brincker2015introduction}. 

Classical modal identification approaches developed within the
operational modal analysis (OMA) framework have been widely studied
over several decades. Methods such as stochastic subspace identification
(SSI), frequency domain decomposition (FDD), and enhanced frequency
domain decomposition (EFDD) offer systematic procedures for extracting
modal parameters from measured response data
\cite{kang2021comparison,worden2020brief,bhowmik2020damping}. Although these methods are
well established and can perform effectively under controlled
conditions, several limitations emerge in real-world structural
applications \cite{ma2023treatment}. In particular, many
frequency-domain techniques still rely on manual or semi-automated
peak-picking procedures, which introduce subjectivity and restrict
scalability for large datasets. The identification of closely spaced or
weakly excited modes remains demanding, and may lead to inaccurate or
unstable estimates \cite{wu2019modal}. Recent developments in online
and recursive modal identification have addressed some of these issues
for real-time applications \cite{bhowmik2024advancements,bhowmik2022real}.
Decomposition-based methods have also improved modal separation and
feature extraction under noisy conditions
\cite{MEHULKUMARR2026119676,bhowmik2020real}. Together, these limitations motivate the
need for more robust, automated, and noise-resilient modal
identification strategies.

In parallel, deep learning approaches have increasingly been explored
for structural analysis and modal identification. Convolutional neural
networks (CNNs), autoencoders, and recurrent architectures have shown
promise in automating feature extraction and improving computational
efficiency for large datasets \cite{alzubaidi2021review}. Physics-
informed and autonomous learning frameworks have further been
investigated to improve robustness and adaptability in SHM
applications \cite{sharma2026autonomous}. Encoder-decoder architectures
such as U-Net \cite{ronneberger2015u} have been explored in related
domains for multi-scale feature extraction and reconstruction,
motivating their integration with graph-based learning frameworks.
Graph neural networks (GNNs) have emerged as a powerful framework for
structural applications because they naturally operate on graph-
structured data and incorporate structural connectivity through message-
passing mechanisms
\cite{kipf2016semi,hamilton2017inductive}. In the context of modal
identification, GNN-based models have demonstrated the ability to learn
from frequency-domain PSD representations while exploiting spatial
relationships between sensors \cite{jian2025using}. Building on these
ideas, variational graph autoencoders (VGAEs)
\cite{kipf2016variational} offer a principled probabilistic extension
that allows structural variability and epistemic uncertainty to be
represented within the latent space, which is directly relevant to the
identification problem considered here.

Despite these advances, important limitations remain in the field of
modal identification. First, many methods treat modal quantities in a
fragmented manner, focusing on either global parameters such as natural
frequencies and damping ratios, or on mode shapes in isolation
\cite{albakri2020estimating}. Unified frameworks that jointly estimate
global modal characteristics and spatially distributed mode shapes
within a single graph-based model are still limited. Second, most
data-driven approaches, including neural network-based models, provide
deterministic predictions or heuristic uncertainty measures, without
offering calibrated predictive distributions or explicitly separating
epistemic and aleatoric sources of uncertainty
\cite{gal2015dropout,lakshminarayanan2016simple}. This limits their
reliability in safety-critical SHM applications. Third, graph-based
methods for structural applications often do not embed fundamental
physical constraints from structural dynamics. In particular, mode
shapes are seldom treated as node-level fields governed by orthogonality
and similarity constraints, leading to physically inconsistent
predictions. Many existing architectures also remain deterministic and
do not adopt variational formulations to capture structural variability
\cite{kingma2013auto}. Finally, current training strategies commonly
optimise for predictive accuracy alone, without explicitly balancing
accuracy, physical consistency, and uncertainty calibration.

To address these limitations, this work proposes a physics-aware
graph-based framework for probabilistic modal identification of
structural systems, evaluating mode-shape accuracy using the Modal
Assurance Criterion (MAC) \cite{allemang2003modal}. The proposed
approach employs a residual U-Net-inspired variational graph
autoencoder (termed UResVGAE), in which the U-Net structure
\cite{ronneberger2015u} enables multi-scale feature aggregation
through encoder--decoder skip connections, improving the reconstruction
of spatially distributed mode shapes. The model learns from PSD-based
structural graphs and enables the joint prediction of natural
frequencies, damping ratios, and spatially distributed mode shapes
within a unified architecture. A dedicated node-level decoding
mechanism reconstructs full-field mode shapes by integrating global
latent representations with local graph features, facilitating improved
representation of higher-order modal characteristics. To promote
physical consistency, the learning process incorporates structural
constraints including mode-shape orthogonality and similarity measures.
A probabilistic framework is introduced to provide calibrated
uncertainty estimates for modal parameters. The effectiveness of the
proposed approach is demonstrated through systematic numerical
experiments under varying noise conditions, indicating strong agreement
in modal parameter estimation and consistent predictive performance.
This suggests that the framework provides a promising first step towards
application in structural health monitoring, with further validation
required on experimental and real-world datasets.

\section{Problem Formulation}
\label{sec:problem_formulation}

In this research, the structural system under consideration is a truss composed of interconnected members and joints that define its geometry and load-bearing behaviour. The structure is represented as a graph $\mathcal{G} = (\mathcal{V}, \mathcal{E})$, where the nodes $\mathcal{V}$ correspond to structural joints and the edges $\mathcal{E}$ represent truss members, thereby encoding the underlying structural connectivity. Each node is associated with a PSD response vector together with its spatial coordinates, capturing the dynamic response and geometric configuration, respectively. The objective is to learn a mapping from this graph-structured representation to the corresponding modal properties of the system.

Formally, given a dataset of graphs, the task is to learn a function $\mathcal{F}$ such that
\begin{equation}
	\mathcal{F}: \mathcal{G} \rightarrow \left( \mathbf{f}, \boldsymbol{\zeta}, \boldsymbol{\Phi} \right),
\end{equation}
where $\mathbf{f} \in \mathbb{R}^{M}$ denotes the natural frequencies, $\boldsymbol{\zeta} \in \mathbb{R}^{M}$ denotes the damping ratios, and $\boldsymbol{\Phi} \in \mathbb{R}^{N \times M}$ denotes the corresponding mode shapes for the first $M$ modes over the $N$ nodes of each graph.

While natural frequencies and damping ratios are global structural properties, mode shapes are spatially distributed quantities defined over the nodes of the graph. Accordingly, for each predicted quantity, the model is required to produce a predictive distribution $p(y \mid \mathcal{G})$, where $y$ denotes the target variable, rather than a deterministic estimate. In particular, the predicted confidence intervals (CI) must be calibrated such that
\begin{equation}
	P\left( y_{\text{true}} \in \text{CI}_{\alpha} \right) \approx \alpha, \quad \forall \alpha \in (0,1).
	\label{eq:calibration}
\end{equation}

Here, $y_{\text{true}}$ denotes the ground-truth value of the parameter to be estimated, and $\alpha$ denotes the prescribed confidence level associated with the corresponding confidence interval.

This calibration requirement, as expressed in Eq.~\eqref{eq:calibration}, is essential for safety-critical decision-making, since mis-calibrated predictions, whether overconfident or under-confident, can lead to unreliable assessment of structural behaviour. Accordingly, the objective of this work is to develop a learning framework that jointly estimates modal properties together with their associated uncertainties \cite{munikoti2022generalframeworkquantifyingaleatoric}, while ensuring both high predictive accuracy and statistically consistent uncertainty quantification. The formulation and evaluation of this framework are presented in the subsequent sections.

\section{Proposed Method}
\label{sec:method}

\begin{figure}[t]
	\centering
	\includegraphics[width=1\linewidth]{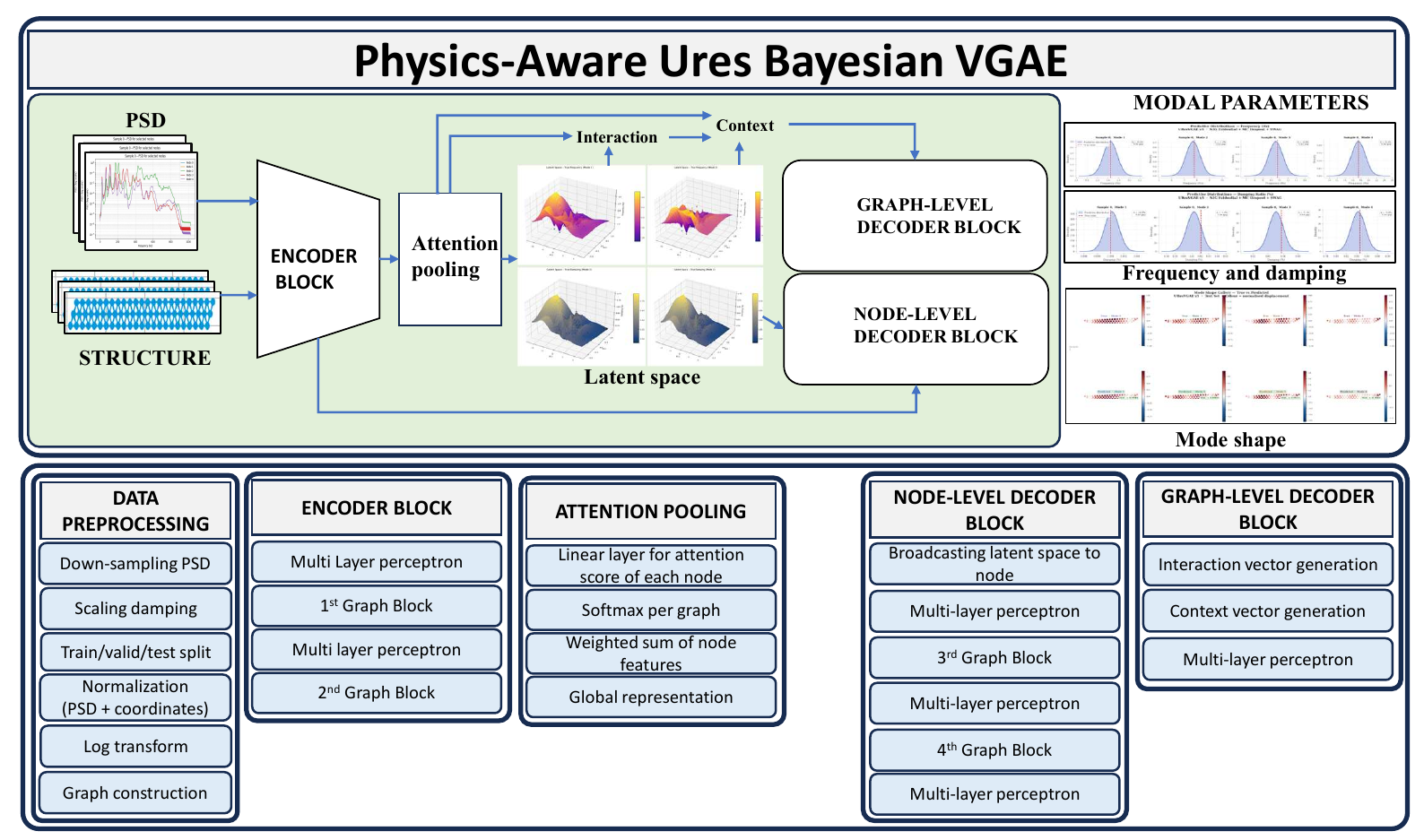}
	\caption{Overview of the proposed physics-aware UResVGAE framework. The figure summarizes the main preprocessing stages and the interaction between local graph features and global latent information.}
	\label{Architecture prelim layout}
\end{figure}

This section presents the proposed U-Net-inspired Residual Variational Graph Autoencoder (UResVGAE) for modal identification from graph-structured PSD data. The framework is designed to learn both graph-level and node-level dynamic characteristics within a unified architecture. For each structural sample, the structure is represented as a graph in which nodes carry spectral and geometric features, while edges encode structural connectivity. On this basis, the model combines residual graph-based encoding, variational latent modelling, and dedicated decoder branches to estimate natural frequencies, damping ratios, and spatially distributed mode shapes. A graphical overview of the proposed framework is provided in Fig.~\ref{Architecture prelim layout}.

A key component of the framework is the global latent representation, which is obtained from pooled graph features and modelled as a stochastic embedding. This latent variable summarises the overall dynamic state of the structure, including information relevant to global modal behaviour that may not be fully captured through purely local neighbourhood aggregation. During decoding, the global representation is broadcast back to the node level and fused with local graph features, allowing the reconstructed mode shapes to remain informed through both structural context and local spatial variation. The following subsections describe the graph representation, architectural components, uncertainty formulation, and training objective in detail.

\subsection{Graph-Based Modal Representation}
\label{sec:4.1}

Each structural sample is represented as a graph, with one graph
corresponding to one truss configuration in the dataset. The full
dataset comprises $2600$ such truss graphs, of which $2000$ are used for
training, $500$ for validation, and $100$ for testing. Within each graph,
nodes represent structural joints and edges represent truss members
derived from the underlying finite-element connectivity. Because the
trusses differ in topology, the resulting graphs are variable-sized,
which allows UResVGAE to operate directly on structures with different
numbers of joints and members without requiring a fixed mesh
representation.

At the node level, the input features consist of the PSD response together with the spatial coordinates of the corresponding joint. In the present implementation, each node carries a $1025$-dimensional PSD vector and its coordinate information, so that both dynamic and geometric characteristics are available to the network. This representation provides the structural input required by UResVGAE to learn graph-level modal quantities and node-level mode-shape fields in a unified manner.

The edge set $E$ is defined from the structural connectivity obtained through the finite-element discretisation, such that an edge exists between two nodes whenever they are connected through a truss member. GraphSAGE is adopted in place of Graph Convolutional Networks (GCNs) and Graph Attention Networks (GATs), following the findings of Jian et al.~\cite{jian2025using}, where GraphSAGE demonstrated superior performance for PSD-based modal identification on variable-topology truss graphs. To implement bidirectional message passing, the graph is treated as undirected, with each edge included in both directions. This representation enables local interactions between connected joints to be captured effectively through neighbourhood aggregation operations.


\subsection{Model architecture}
\label{sec:4.2}

UResVGAE adopts a residual graph-based encoder--decoder architecture
inspired by the U-Net paradigm \cite{ronneberger2015u}, designed to
capture both global structural dynamics and fine-grained spatial
information. The architecture consists of a spectral encoder for
compressing node-wise PSD features, residual GraphSAGE blocks for
propagating structural information, a variational latent space
\cite{kipf2016variational,kingma2013auto} for capturing graph-level
dynamics, and separate decoder branches for predicting global modal
parameters and node-level mode shapes with associated uncertainty

\subsubsection{Spectral Feature Encoding}

The PSD associated with each node is represented as a high-dimensional
feature vector containing modal information together with environmental
noise. Using this raw feature directly can hinder learning, since
redundant spectral content and measurement noise are propagated through
subsequent graph operations. A spectral feature encoding step is
therefore introduced to map the input PSD to a lower-dimensional latent
representation before graph-based message passing. This stage acts as
a reduced-order spectral projection that compresses the response data
while retaining the dominant modal characteristics required for
subsequent prediction.

This transformation is implemented through a multi-layer perceptron
(MLP), which learns a nonlinear projection from the original PSD space
to a compact feature space. Beyond dimensionality reduction, the
encoding stage reorganises the spectral information into a form more
suitable for graph learning, thereby improving robustness to redundant
frequency content and measurement noise. As a result, the encoded node
features provide a compact and informative representation of the
structural response that can be propagated more effectively through the
subsequent GraphSAGE layers. For a node $i$ with PSD vector
$\mathbf{x}_i \in \mathbb{R}^F$, the encoded feature
$\mathbf{h}_i^{(0)} \in \mathbb{R}^D$ is obtained as:
\begin{equation}
	\mathbf{h}_i^{(0)} = f_{\text{MLP}}(\mathbf{x}_i),
\end{equation}
where $f_{\text{MLP}}(\cdot)$ denotes the nonlinear mapping defined
through the spectral encoder. This encoded representation serves as the
initial node feature for subsequent graph-based learning.

\subsubsection{Graph block}

The encoded features are then propagated through a sequence of residual graph blocks based on the GraphSAGE operator \cite{hamilton2018inductiverepresentationlearninglarge}, which performs neighbourhood aggregation to capture local structural interactions.

Formally, let $\mathbf{h}_i^{(l)} \in \mathbb{R}^d$ denote the feature vector of node $i$ at layer $l$. The GraphSAGE update can be expressed as:
\begin{equation}
	\mathbf{h}_i^{(l+1)} = \sigma \left( \mathbf{W}^{(l)} \cdot
	\text{AGG} \left( \{ \mathbf{h}_i^{(l)} \} \cup
	\{ \mathbf{h}_j^{(l)} \mid j \in \mathcal{N}(i) \} \right) \right),
\end{equation}
where $\mathcal{N}(i)$ denotes the set of neighbouring nodes of node $i$, $\text{AGG}(\cdot)$ represents an aggregation function, taken here as mean aggregation, $\mathbf{W}^{(l)}$ is a learnable weight matrix, and $\sigma(\cdot)$ is a nonlinear activation function.

To improve training stability and enable deeper feature representations, residual connections are incorporated at each layer \cite{chen2022resgraphnet}. The residual formulation is given by:
\begin{equation}
	\mathbf{h}_i^{(l+1)} = \mathbf{h}_i^{(l)} +
	\sigma \left( \mathbf{W}^{(l)} \cdot
	\text{AGG} \left( \{ \mathbf{h}_i^{(l)} \} \cup
	\{ \mathbf{h}_j^{(l)} \mid j \in \mathcal{N}(i) \} \right) \right).
\end{equation}

An intermediate feature projection is introduced before the second GraphSAGE block to map the node embeddings from $\mathbb{R}^{d}$ to a higher-dimensional space $\mathbb{R}^{d_2}$. Formally, if $\mathbf{H}^{(1)} \in \mathbb{R}^{N \times d}$ denotes the output of the first graph block, namely the matrix collecting the feature vectors of all nodes at that layer, the transformed features are written as
\[
\tilde{\mathbf{H}}^{(1)} = \mathbf{H}^{(1)} \mathbf{P}, \qquad \mathbf{P} \in \mathbb{R}^{d \times d_2}, \quad d_2 > d.
\]
This expansion does not enlarge the graph receptive field, but it provides a richer channel space within which the second GraphSAGE layer can encode the aggregated neighbourhood information \cite{song2021networkgraphneuralnetwork}. Consequently, after two message-passing layers, each node representation is learned from successively aggregated one-hop interactions, yielding an effective two-hop receptive field in a more expressive feature space. In the proposed model, two residual GraphSAGE blocks are employed, producing the node-level embeddings $\mathbf{H}^{(1)}$ and $\mathbf{H}^{(2)}$.

\subsubsection{Attention pooling block}

To derive a compact graph-level representation for each truss sample,
the relative contribution of individual nodes must be considered, since
global structural behaviour emerges from the collective interaction of
local components. Following the graph block, an attention pooling
mechanism is introduced to identify and aggregate the most informative
node features. This pooled representation preserves the structural
information required for predicting global modal properties such as
natural frequencies and damping ratios, and is formulated in an
attention-based manner following \cite{Itoh_2022}.

Let $\mathbf{h}_i \in \mathbb{R}^d$ denote the feature vector of node $i$. An attention score $a_i$ is first computed for each node using a learnable linear projection:
\begin{equation}
	a_i = \mathbf{q}^\top \mathbf{h}_i,
\end{equation}
where $\mathbf{q} \in \mathbb{R}^d$ is a learnable parameter vector.

The attention weights are then obtained by normalizing these scores across all nodes using a softmax function:
\begin{equation}
	w_i = \frac{\exp(a_i)}{\sum_{j \in \mathcal{V}} \exp(a_j)},
\end{equation}
where $\mathcal{V}$ denotes the set of nodes in the graph.

Finally, the graph-level representation $\mathbf{h}_g$ is computed as a weighted sum of node features:
\begin{equation}
	\mathbf{h}_g = \sum_{i \in \mathcal{V}} w_i \mathbf{h}_i.
\end{equation}

\subsubsection{Variational Latent Space Formulation}

To account for structural variability and predictive uncertainty, a
variational latent space is introduced following the VGAE framework
\cite{kipf2016variational,kingma2013auto}. The graph-level embedding
$\mathbf{h}_g \in \mathbb{R}^d$ obtained from attention pooling is
mapped to a probabilistic latent space parameterised through a Gaussian
distribution, enabling structural variability and uncertainty in the
learned global dynamics to be represented explicitly.

Specifically, the encoder predicts the mean and log-variance of the latent distribution as
\begin{equation}
	\boldsymbol{\mu} = \mathbf{W}_{\mu} \mathbf{h}_g,
	\quad
	\log \boldsymbol{\sigma}^2 = \mathbf{W}_{\sigma} \mathbf{h}_g,
\end{equation}
where $\boldsymbol{\mu}, \log \boldsymbol{\sigma}^2 \in \mathbb{R}^{d_z}$ are the latent mean and log-variance vectors, and $\mathbf{W}_{\mu}$ and $\mathbf{W}_{\sigma}$ are learnable projection matrices.

The latent variable $\mathbf{z}$ is then sampled using the reparameterization trick:
\begin{equation}
	\mathbf{z} = \boldsymbol{\mu} + \mathbf{s} \odot \boldsymbol{\epsilon},
	\quad
	\boldsymbol{\epsilon} \sim \mathcal{N}(\mathbf{0}, \mathbf{I}),
\end{equation}
where $\mathbf{s} = \exp\left(\frac{1}{2}\log \boldsymbol{\sigma}^2\right)$, $\odot$ denotes element-wise multiplication, and $\boldsymbol{\epsilon}$ is sampled from a standard normal distribution. This formulation enables gradient-based optimisation through the expression of the stochastic sampling step as a deterministic transformation of a random noise variable.

To regularise the latent space, the learned distribution is constrained to remain close to a standard normal prior $\mathcal{N}(\mathbf{0}, \mathbf{I})$ through the Kullback--Leibler (KL) divergence \cite{ahn2021variational}:
\begin{equation}
	\mathcal{L}_{\text{KL}} = -\frac{1}{2} \sum_{i=1}^{d_z}
	\left(1 + \log \sigma_i^2 - \mu_i^2 - \sigma_i^2 \right).
\end{equation}

The resulting latent embedding captures graph-level structural variation together with uncertainty in the learned representation, which improves the ability of the model to generalise to unseen structural configurations.

\subsubsection{Node-Level Decoder for Mode Shape Reconstruction}

The node-level decoder is designed to reconstruct spatially resolved mode shapes through the integration of global latent information with local structural features. This is achieved through a sequence of operations involving latent broadcasting, feature fusion, residual graph decoding, and deep skip connections.

\paragraph{a. Latent Broadcasting}
The latent vector $\mathbf{z} \in \mathbb{R}^{d_z}$, obtained from the variational latent space, encodes graph-level structural characteristics. To incorporate this information into node-level prediction, the latent vector is broadcast to all nodes in the graph. For a graph with $N$ nodes, this operation is expressed as:
\begin{equation}
	\mathbf{Z} = \text{repeat}(\mathbf{z}) \in \mathbb{R}^{N \times d_z},
\end{equation}
where each node receives the same latent embedding.

\paragraph{b. Latent Feature Projection and Fusion}

To ensure compatibility with the node-feature dimensions, the broadcast latent vector is first projected into the feature space:
\begin{equation}
	\tilde{\mathbf{Z}} = \mathbf{W}_z \mathbf{Z},
\end{equation}
where $\mathbf{W}_z$ is a learnable projection matrix.

The projected latent features are then concatenated with the encoder output $\mathbf{H}^{(2)}$:
\begin{equation}
	\mathbf{H}^{(c)} = \left[\mathbf{H}^{(2)} \mid \tilde{\mathbf{Z}}\right],
\end{equation}
where $\mid$ denotes feature-wise concatenation. This fusion allows each node to access both local structural information and global contextual information simultaneously. Here, $\mathbf{H}^{(2)}$ encodes node-level dynamic and neighbourhood information of the truss, whereas $\tilde{\mathbf{Z}}$ provides a graph-level summary of the overall modal behaviour. The concatenated features are then processed through a MLP, which learns a joint feature representation and projects it back to a lower-dimensional space.

\paragraph{c. Residual Graph-Based Decoding}

The fused features are processed through a sequence of residual graph convolutional layers to refine the node representations:
\begin{equation}
	\mathbf{H}^{(l+1)} = \mathbf{H}^{(l)} +
	\sigma \left( \mathbf{W}^{(l)} \cdot
	\text{AGG} \left( \mathbf{H}^{(l)} \right) \right),
\end{equation}
where $\text{AGG}(\cdot)$ denotes neighbourhood aggregation and $\sigma(\cdot)$ is a nonlinear activation function.

These layers propagate both local and global information across the graph, enabling the model to reconstruct spatially coherent deformation patterns. In the present decoder, two GraphSAGE blocks are employed, producing the intermediate representations $\mathbf{H}^{(3)}$ and $\mathbf{H}^{(4)}$. An additional MLP is introduced between these blocks to reduce the feature dimension. The overall information flow of this node-level decoding process is illustrated in Fig.~\ref{Architecture prelim layout}.

\paragraph{d. Deep Skip Connections}

To preserve fine-grained spatial details, skip connections are
introduced between the early encoder features $\mathbf{H}^{(1)}$ and
the decoder features. This connection enables the decoder to reuse
early node-level features retaining higher-resolution information about
the local structural response, which may otherwise be weakened after
latent compression and successive graph aggregation steps:
\begin{equation}
	\mathbf{H}^{(skip)} = \left[ \mathbf{H}^{(4)} \mid \mathbf{H}^{(1)} \right].
\end{equation}
This fusion restores high-resolution local information that may be
lost during encoding and latent compression, which is particularly
important for accurate reconstruction of higher-order mode shapes. The
fused representation is then passed through a MLP to learn the
interaction between the concatenated features and reduce the feature
dimension, yielding $\mathbf{H}^{(final)}$.

\paragraph{e. Mode Shape Prediction}

The refined node embeddings are mapped to the mode-shape outputs
through a linear projection:
\begin{equation}
	\boldsymbol{\Phi} = \mathbf{H}^{(final)} \mathbf{W}_{\phi},
\end{equation}
where $\boldsymbol{\Phi} \in \mathbb{R}^{N \times M}$ represents the
predicted mode shapes for $M$ modes across $N$ nodes, and
$\mathbf{W}_{\phi}$ is a learnable weight matrix. This output provides
a full-field spatial representation of structural deformation patterns,
enabling accurate estimation of mode shapes.

\subsubsection{Graph-Level Decoder for Modal Parameter Prediction}

The graph-level decoder is responsible for predicting the global modal parameters, namely natural frequencies and damping ratios, together with their associated uncertainty. This is achieved through the construction of a context representation that integrates graph-level structural information with the stochastic latent embedding.

\paragraph{a. Context Vector Construction}

The decoder operates on a graph-level feature vector formed through the combination of the pooled representation $\mathbf{h}_g \in \mathbb{R}^{d}$ and the latent vector $\mathbf{z} \in \mathbb{R}^{d_z}$. To enhance the interaction between these components, an interaction-based transformation is introduced:
\begin{equation}
	\mathbf{c} = \left[ \mathbf{h}_g \mid \mathbf{z} \mid \left( \mathbf{W}_i \mathbf{h}_g \odot \mathbf{z} \right) \right],
\end{equation}
where $\mathbf{W}_i$ is a learnable projection matrix, $\odot$ denotes element-wise multiplication, and $\mid$ represents concatenation. This formulation captures not only the individual contributions of the graph-level and latent features, but also their interaction, thereby yielding a more expressive contextual representation.

\paragraph{b. Evidential Regression Head}

The context vector $\mathbf{c}$ is passed through a deep neural network to predict the parameters of a Normal--Inverse-Gamma (NIG) distribution \cite{amini2020deep,ma2021trustworthy} for each modal quantity:
\begin{equation}
	(\gamma, \nu, \alpha, \beta) = f_{\text{head}}(\mathbf{c}),
\end{equation}
\textbf{where $\gamma \in \mathbb{R}^{M}$ denotes the predicted mean, $\nu \in \mathbb{R}^{M}_{>0}$ is the evidence or virtual observation parameter, and $\alpha, \beta \in \mathbb{R}^{M}_{>0}$ are the shape and scale parameters of the Normal--Inverse-Gamma distribution, respectively. Here, $M$ denotes the number of predicted modes, and $f_{\text{head}}(\cdot)$ denotes a multi-layer perceptron with nonlinear activations.}

To ensure valid uncertainty estimates, constraints are imposed on these parameters:
\begin{equation}
	\nu = \text{softplus}(\tilde{\nu}) + \epsilon, \quad
	\alpha = \text{softplus}(\tilde{\alpha}) + 1 + \epsilon, \quad
	\beta = \text{softplus}(\tilde{\beta}) + \epsilon,
\end{equation}
where $\tilde{\nu}$, $\tilde{\alpha}$, and $\tilde{\beta}$ are unconstrained outputs of the network, and $\epsilon$ is a small constant introduced for numerical stability.

\paragraph{c. Predictive Distribution}

Given the predicted NIG parameters, the target variable $y$ corresponding to frequency or damping follows a Student-$t$ predictive distribution:
\begin{equation}
	p(y \mid \gamma, \nu, \alpha, \beta) = \text{Student-}t\left(y \mid \gamma, \frac{\beta(1+\nu)}{\nu(\alpha-1)}, 2\alpha \right).
	\label{eq:21}
\end{equation}

The predictive mean is given by:
\begin{equation}
	\mathbb{E}[y] = \gamma,
\end{equation}
while the predictive variance is:
\begin{equation}
	\sigma^2 = \frac{\beta}{\alpha - 1} \left(1 + \frac{1}{\nu}\right).
	\label{variance}
\end{equation}

\paragraph{d. Dual-Head Prediction}

An evidential head is used to predict the frequency-related NIG parameters:
\begin{equation}
	(\gamma_f, \nu_f, \alpha_f, \beta_f) = f_{\text{freq}}(\mathbf{c}),
\end{equation}
where $\gamma_f$, $\nu_f$, $\alpha_f$, and $\beta_f$ define the Normal--Inverse-Gamma distribution associated with the predicted natural frequencies. An analogous evidential head is employed for damping-ratio prediction, yielding the corresponding parameters $(\gamma_\zeta, \nu_\zeta, \alpha_\zeta, \beta_\zeta)$.

This separation allows the model to learn task-specific representations for each modal quantity while sharing a common contextual embedding.

\subsection{Uncertainty Quantification Framework}
\label{sec:4.3}
A primary objective of this work is to provide not only accurate
predictions of modal parameters but also reliable and calibrated
uncertainty estimates. To this end, a hybrid uncertainty quantification
framework is proposed, integrating evidential deep learning with
Bayesian approximation techniques. This combination enables the model
to capture both data-driven and model-related uncertainty in a
complementary manner, producing calibrated predictive distributions
suited for safety-critical structural applications.

\subsubsection{Evidential Uncertainty Modelling}

The primary source of uncertainty estimation in the proposed framework is based on Evidential Deep Learning (EDL) \cite{amini2020deep,gao2025comprehensive}. Instead of producing a standard point estimate, the model outputs the parameters of a Normal--Inverse-Gamma (NIG) distribution for each target variable:
\begin{equation}
	\boldsymbol{m} = (\gamma, \nu, \alpha, \beta),
\end{equation}
where $\gamma$ denotes the predicted mean, $\nu > 0$ is the virtual observation count or evidence parameter, and $\alpha > 1$ and $\beta > 0$ characterise the shape and scale of the underlying variance.

The predictive distribution of the target variable $y$, obtained
through marginalisation over the NIG parameters, follows the
Student-$t$ form given in Eq.~\ref{eq:21}. The evidential output
therefore provides both the predictive mean and an analytical
decomposition of uncertainty into epistemic and aleatoric components.

From this distribution, the expected value is given by:
\begin{equation}
	\mathbb{E}[y] = \gamma.
\end{equation}

The predictive variance decomposes analytically into epistemic and
aleatoric components. Epistemic uncertainty, denoted
$\sigma_{\text{epis}}^2$, represents the lack of model knowledge
arising from limited or insufficiently informative training data:
\begin{equation}
	\sigma_{\text{epis}}^2 = \frac{\beta}{\nu(\alpha - 1)}.
\end{equation}

Aleatoric uncertainty, denoted $\sigma_{\text{alea}}^2$, captures the
inherent stochastic variability in the observations, such as sensor
noise in the measured truss response:
\begin{equation}
	\sigma_{\text{alea}}^2 = \frac{\beta}{\alpha - 1}.
\end{equation}

The corresponding total predictive variance is therefore expressed as:
\begin{equation}
	\sigma_{\text{total}}^2 = \frac{\beta}{\alpha - 1} \left(1 + \frac{1}{\nu} \right).
\end{equation}

A practical advantage of this formulation -- consistent with
Eq.~\eqref{variance} -- is that it enables simultaneous estimation of
the predictive mean and uncertainty within a single forward pass,
avoiding the computational overhead associated with Monte Carlo dropout
or deep ensembles while retaining analytical tractability.

\subsubsection{Bayesian Inference via SWAG and MC Dropout}

Evidential learning provides the primary uncertainty estimate within
a single forward pass. In addition, Bayesian approximation techniques
are employed at inference time to assess uncertainty arising from the
model parameters. Accordingly, MC dropout \cite{gal2015dropout} and
Stochastic Weight Averaging Gaussian (SWAG) \cite{maddox2019simple}
are used as supplementary sampling-based mechanisms rather than as the
principal uncertainty estimator.

\paragraph{a. Monte Carlo Dropout}
MC dropout approximates Bayesian inference through stochastic forward
passes with dropout activated during inference. Let $\hat{y}^{(t)}$
denote the prediction obtained from the $t^{\text{th}}$ stochastic
forward pass. For $T$ such passes, the predictive mean is estimated as:
\begin{equation}
	\mathbb{E}[\hat{y}^{(t)}] \approx \frac{1}{T} \sum_{t=1}^{T} \hat{y}^{(t)}.
\end{equation}

The corresponding uncertainty is quantified through the variance across
these stochastic samples:
\begin{equation}
	\sigma_{\text{MC}}^2 \approx \frac{1}{T} \sum_{t=1}^{T} \left(\hat{y}^{(t)} - \mathbb{E}[y]\right)^2.
\end{equation}

\paragraph{b. Stochastic Weight Averaging Gaussian (SWAG)}
SWAG approximates the posterior distribution over network parameters
by fitting a Gaussian distribution to model weights collected during
training \cite{maddox2019simple}. Let $\boldsymbol{\theta}^{(k)}$
denote the model parameters at iteration $k$. The mean and covariance
of the approximate posterior are computed as:
\begin{equation}
	\boldsymbol{\mu}_{\theta} = \frac{1}{K} \sum_{k=1}^{K} \boldsymbol{\theta}^{(k)},
\end{equation}
\begin{equation}
	\boldsymbol{\Sigma}_{\theta} = \frac{1}{K-1} \sum_{k=1}^{K} \left(\boldsymbol{\theta}^{(k)} - \boldsymbol{\mu}_{\theta}\right)^2,
\end{equation}
where $K$ is the number of collected models.

At inference, weights are sampled from the Gaussian posterior:
\begin{equation}
	\boldsymbol{\theta}^{(s)} \sim \mathcal{N}(\boldsymbol{\mu}_{\theta}, \boldsymbol{\Sigma}_{\theta}),
\end{equation}
and predictions are obtained for each sampled model:
\begin{equation}
	\hat{y}^{(s)} = f_{\boldsymbol{\theta}^{(s)}}(x).
\end{equation}

The predictive mean and variance are then estimated as:
\begin{equation}
	\mathbb{E}[y] \approx \frac{1}{S} \sum_{s=1}^{S} \hat{y}^{(s)}, \quad
	\sigma_{\text{SWAG}}^2 \approx \frac{1}{S} \sum_{s=1}^{S} \left(\hat{y}^{(s)} - \mathbb{E}[y]\right)^2.
\end{equation}

The overall predictive distribution is obtained by combining evidential
uncertainty with stochastic sampling from MC dropout and SWAG. In this
way, total uncertainty reflects both data-driven uncertainty (evidential
learning) and model-parameter uncertainty (stochastic inference),
resulting in a more robust and better-calibrated predictive estimate.

\subsection{Training Objective}
\label{sec:4.4}

To model both predictive accuracy and uncertainty, an evidential
regression loss based on the NIG distribution is employed for frequency
and damping estimation. For a target variable $y$ with predicted
parameters $(\gamma, \nu, \alpha, \beta)$, the negative log-likelihood
($\mathcal{L}_{\text{NLL}}$) is given by
\begin{equation}
	\mathcal{L}_{\text{NLL}} =
	\frac{1}{2}\log\left(\frac{\pi}{\nu}\right)
	- \alpha \log\left(2\beta(1+\nu)\right)
	+ \left(\alpha + \frac{1}{2}\right)\log\left((y-\gamma)^2 \nu + 2\beta(1+\nu)\right)
	+ \log \Gamma(\alpha)
	- \log \Gamma\left(\alpha + \frac{1}{2}\right)
\end{equation}
where $\Gamma(\cdot)$ denotes the Gamma function.

To discourage overconfident predictions, an evidential regularization term is introduced:
\begin{equation}
	\mathcal{L}_{\text{reg}} = \left| y - \gamma \right| (2\nu + \alpha)
\end{equation}

The final evidential loss for frequency and damping estimation is defined as
\begin{equation}
	\mathcal{L}_{\text{freq}} = \mathcal{L}_{\text{zeta}} =
	\mathcal{L}_{\text{NLL}} + \lambda_{\text{evi}} \mathcal{L}_{\text{reg}}.
\end{equation}

To improve calibration, a loss based on the Continuous Ranked
Probability Score (CRPS) is incorporated
\cite{zamo2018estimation,guo2017calibration}. For a Gaussian predictive
distribution with mean $\mu$ and standard deviation $\sigma$, the CRPS
is defined as
\begin{equation}
	\mathcal{L}_{\text{CRPS}} =
	\sigma \left[
	z \left(2\Phi(z) - 1\right)
	+ 2\phi(z)
	- \frac{1}{\sqrt{\pi}}
	\right]
\end{equation}
where $z = \frac{y - \mu}{\sigma}$, and $\Phi(\cdot)$ and $\phi(\cdot)$ denote the cumulative distribution function and probability density function of the standard normal distribution, respectively.

For mode-shape reconstruction, the MAC is used to measure the similarity between predicted mode shapes $\boldsymbol{\phi}_p$ and theoretical mode shapes $\boldsymbol{\phi}_t$:
\begin{equation}
	\text{MAC} =
	\frac{\left|\boldsymbol{\phi}_p^\top \boldsymbol{\phi}_t\right|^2}
	{\left(\boldsymbol{\phi}_p^\top \boldsymbol{\phi}_p\right)
		\left(\boldsymbol{\phi}_t^\top \boldsymbol{\phi}_t\right)}.
\end{equation}

The corresponding loss is defined as
\begin{equation}
	\mathcal{L}_{\text{MAC}} = 1 - \text{MAC}.
\end{equation}

To enforce physical consistency, an orthogonality constraint is applied across the predicted mode shapes. The Gram matrix is defined as $\mathbf{G} = \boldsymbol{\Phi}^\top \boldsymbol{\Phi}$, and the orthogonality loss is given by
\begin{equation}
	\mathcal{L}_{\text{ortho}} = \left| \mathbf{G} - \mathbf{I} \right|_1,
\end{equation}
where $\mathbf{I}$ is the identity matrix and $|\cdot|_1$ denotes the element-wise $\ell_1$ norm.

Finally, the latent space is regularized using the KL divergence:
\begin{equation}
	\mathcal{L}_{\text{KL}} =
	-\frac{1}{2} \sum \left(1 + \log \sigma^2 - \mu^2 - \sigma^2 \right),
\end{equation}
which constrains the learned latent distribution to remain close to a standard Gaussian prior.

The overall training objective is defined as
\begin{equation}
	\mathcal{L} = \lambda_f \mathcal{L}_{\text{freq}} 
	+ \lambda_\zeta \mathcal{L}_{\text{zeta}} 
	+ \lambda_{\text{CRPS}} \mathcal{L}_{\text{CRPS}} 
	+ \lambda_{\phi} \mathcal{L}_{\text{MAC}} 
	+ \lambda_{\text{ortho}} \mathcal{L}_{\text{ortho}} 
	+ \beta \mathcal{L}_{\text{KL}}
	\label{eq:overall_loss}
\end{equation}
where $\lambda_f$ and $\lambda_\zeta$ are the weights associated with the evidential regression losses for natural frequency and damping ratio, respectively, $\lambda_{\text{CRPS}}$ is the weight assigned to the calibration loss, $\lambda_{\phi}$ is the weight for the mode-shape reconstruction loss, $\lambda_{\text{ortho}}$ is the weight for the orthogonality constraint, and $\beta$ is the weight associated with the KL regularization term. This formulation enables the model to achieve accurate predictions while maintaining physically consistent outputs and well-calibrated uncertainty estimates.

\subsection{Training Strategy}
\label{sec:4.5}

To stabilise optimisation, the training process is structured into three
phases with progressively increasing objective complexity. The strategy
is designed to first establish reliable graph-level modal
representations, then improve node-level mode-shape reconstruction,
and finally activate the full uncertainty-aware and physics-constrained
objective. Algorithm~\ref{alg:training_strategy} summarises the phase-
wise training strategy, and the main components are outlined below:

\begin{itemize}
	\item \textbf{Phase 1: Global modal learning.} 
	The model is first trained using only the evidential regression losses for natural frequencies and damping ratios. The mode-shape reconstruction loss is excluded at this stage so that the graph-level decoder can learn meaningful global representations without the added difficulty of node-level spatial reconstruction.
	
	\item \textbf{Phase 2: Mode-shape learning.} 
	The mode-shape reconstruction loss is then introduced to improve the MAC values of the predicted mode shapes. Greater emphasis is placed on higher-order modes, since they are more sensitive to local structural variation and are typically more difficult to reconstruct accurately.
	
	\item \textbf{Phase 3: Full objective activation.} 
	In the final phase, the complete loss function is activated, including the orthogonality constraint and calibration-related terms. During this stage, SWAG is used to approximate the posterior distribution over the model parameters, while MC dropout is applied at inference time through multiple stochastic forward passes to provide an additional assessment of parameter uncertainty.
\end{itemize}

In addition, warm-up scheduling is applied to selected loss terms. The KL-divergence weight is increased gradually to reduce the risk of posterior collapse in the latent space, while the evidential regularization term is introduced progressively to avoid over-penalising prediction errors during the early stages of training. The CRPS-based calibration loss is activated only after the model has learned sufficiently accurate predictive means. Gradient clipping is further employed to improve numerical stability, and a cosine-annealing learning-rate schedule with warm restarts is used to support convergence.

\begin{algorithm}[H]
	\caption{Phase-wise training strategy of the proposed UResVGAE framework}
	\label{alg:training_strategy}
	\begin{algorithmic}[1]
		\State Initialize model, optimizer, scheduler, and loss weights
		\State Warm up $\beta$ and evidential regularization; delay $\mathcal{L}_{\text{CRPS}}$
		
		\Statex
		\State \textbf{Phase 1: Global modal learning}
		\For{each epoch}
		\State Optimize $\mathcal{L}_{\text{freq}} + \mathcal{L}_{\zeta}$
		\EndFor
		
		\Statex
		\State \textbf{Phase 2: Mode-shape learning}
		\For{each epoch}
		\State Optimize $\mathcal{L}_{\text{freq}} + \mathcal{L}_{\zeta} + \mathcal{L}_{\text{MAC}}$
		\EndFor
		
		\Statex
		\State \textbf{Phase 3: Full objective}
		\For{each epoch}
		\State Optimize the full loss in Eq.~(\ref{eq:overall_loss})
		\State Update warm-up terms and activate $\mathcal{L}_{\text{CRPS}}$ after stabilisation
		\EndFor
		
		\Statex
		\State \textbf{Inference}
		\State Use evidential prediction; apply MC dropout and SWAG for supplementary uncertainty estimation
	\end{algorithmic}
\end{algorithm}

\section{Experimental Setup}
\label{sec:experimental_setup}

A dataset of 2600 truss structures within a trapezoidal domain is
generated using Delaunay triangulation and simulated through finite
element analysis. Of these, 2000 trusses are used for training, 500
for validation, and 100 for testing, with modal parameters obtained
through eigenvalue analysis \cite{jian2025using}. Each truss contains
a variable number of nodes, and each node is associated with a
structural response represented in the form of PSD, computed using
Welch's method \cite{welch1967use}. Prior to training, all input
features, including PSDs and node coordinates, are normalised to zero
mean and unit standard deviation to ensure stable optimisation. To
reduce computational complexity, the PSD signals are downsampled from
1024 to 512 frequency bins along the frequency axis. Natural
frequencies and damping ratios are log-transformed to balance the
scale between lower and higher modes, thereby stabilising the loss
function and reducing the risk of gradient explosion.

The model is implemented using PyTorch and the Deep Graph Library
(DGL). The AdamW optimiser is employed with separate learning rates
for the backbone and prediction heads. Mini-batch training with
gradient accumulation is used to simulate larger batch sizes, while
gradient clipping is applied to ensure stable backpropagation. Model
performance is evaluated using the MAC \cite{allemang2003modal}
for mode shapes and relative error metrics for natural frequencies and
damping ratios. To assess the quality of uncertainty estimation,
calibration metrics are also considered. The Expected Calibration Error
(ECE) \cite{guo2017calibration,pavlovic2025understanding} is used to
quantify the discrepancy between predicted confidence levels and
observed coverage. Empirical coverage probabilities are also evaluated
across multiple confidence intervals to assess calibration performance.
The final model is selected on the basis of validation performance and
subsequently evaluated on the test dataset.

\section{Results}
\label{sec:results}

The predictive performance of the proposed framework is assessed
through a combination of quantitative metrics and qualitative
visualisations. The evaluation examines error distributions,
reconstruction fidelity, and consistency across different modal
quantities and test conditions, providing a more complete assessment
of model accuracy, robustness, and overall reliability than aggregate
statistics alone.

\subsection{Statistical Characteristics of Modal Parameter Estimation}

\begin{figure}[h]
	\centering
	\includegraphics[width=1\linewidth]{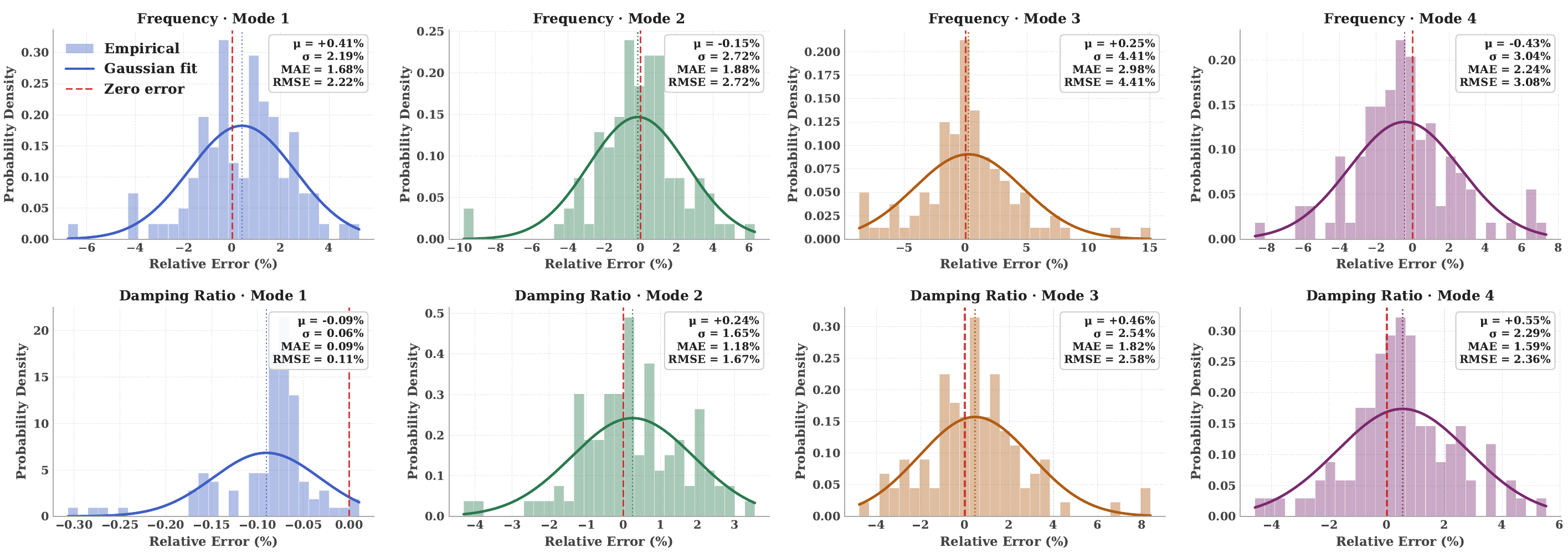}
	\caption{Signed relative error distributions for natural frequency and damping ratio predictions across all test samples. The top row shows the frequency errors for Modes 1--4, and the bottom row shows the corresponding damping ratio errors.}
	\label{fig:error_distribution}
\end{figure}

The predictive performance of the proposed model is first examined
through a statistical analysis of errors across all test samples and
modes. Fig.~\ref{fig:error_distribution} presents the signed relative
error distributions for natural-frequency and damping-ratio estimation
across all four modes. For natural frequencies, the distributions
remain tightly centred around zero, indicating low prediction bias.
Most estimates fall within a narrow error range, although the spread
increases gradually for higher modes, which is consistent with their
greater sensitivity and reduced observability. A similar trend is
observed for damping-ratio estimation, although the damping-error
distributions are noticeably wider than those of the corresponding
frequency predictions. This behaviour is consistent with the greater
sensitivity of damping estimation to noise and spectral resolution.
Nevertheless, the damping predictions remain approximately centred,
suggesting that the model does not exhibit strong systematic bias.

\subsection{Mode-Shape Reconstruction Performance}

\begin{figure}[t]
	\centering
	\includegraphics[width=1\linewidth]{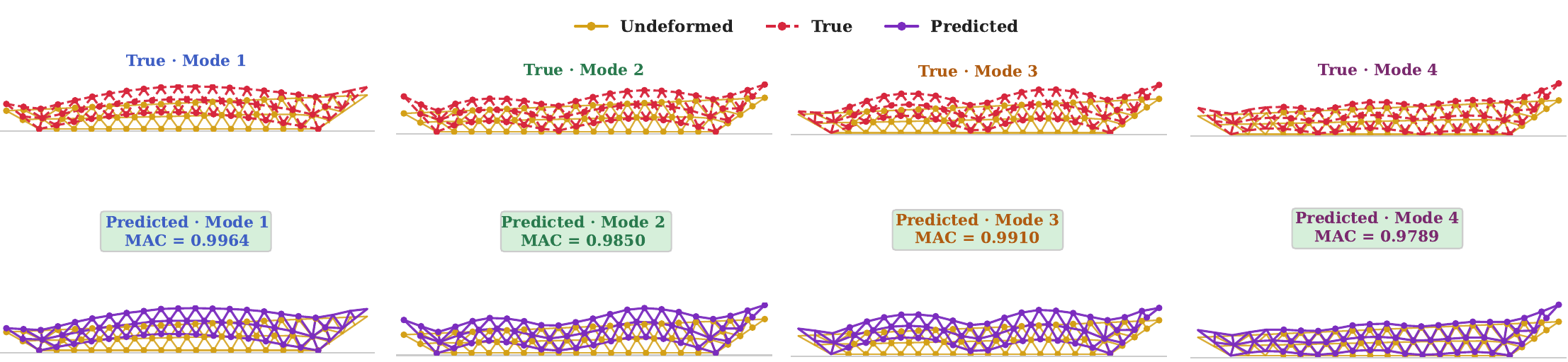}
	\caption{Comparison of true and predicted mode shapes for a representative test sample. The top row shows the reference mode shapes for Modes 1--4, and the bottom row shows the corresponding model predictions. The associated MAC values indicate strong agreement between the predicted and reference spatial patterns across all modes.}
	\label{fig:mode_shape_gallery}
\end{figure}

The quality of the predicted mode shapes is examined through both sample-level and dataset-level comparisons. Fig.~\ref{fig:mode_shape_gallery} shows the true and predicted mode shapes for a representative test sample, demonstrating close agreement in the spatial deformation patterns across all four modes. This visual consistency is supported through the corresponding MAC values, which remain high for each mode.

\begin{figure}[H]
	\centering
	\includegraphics[width=0.85\linewidth]{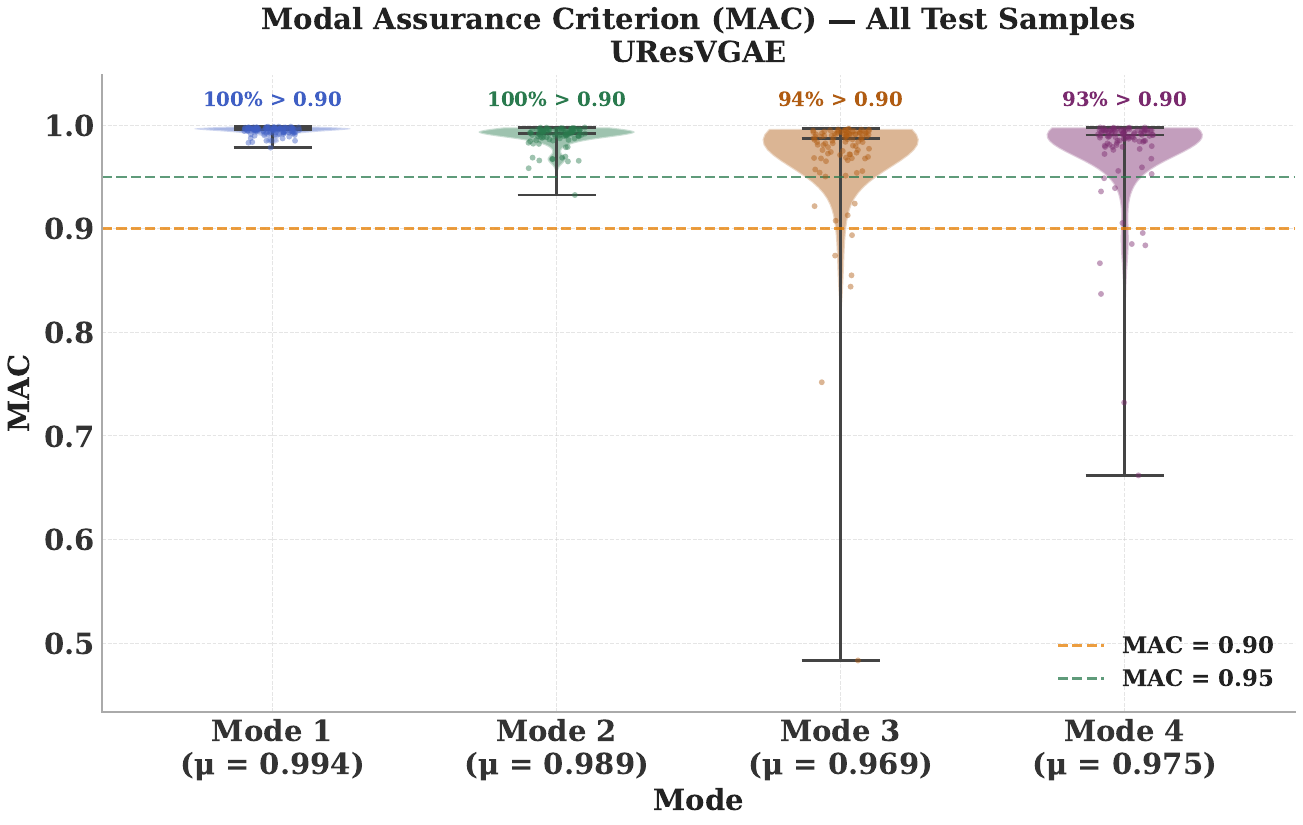}
	\caption{Distribution of MAC values across all test samples for Modes 1--4. The violin plots illustrate the spread of mode-shape agreement, while the annotated mean values and threshold lines at MAC = 0.90 and MAC = 0.95 provide a direct indication of reconstruction quality. The results show consistently high MAC values, with slightly increased variability for the higher modes.}
	\label{fig:mac_violin}
\end{figure}

To assess mode-shape performance over the full test set, Fig.~\ref{fig:mac_violin} presents the distribution of MAC values for all modes. The majority of predictions remain close to unity, confirming strong similarity between predicted and reference mode shapes. The mean MAC values remain high for all modes, with only moderate degradation observed for the higher modes, where reconstruction becomes more challenging. Taken together, Figs.~\ref{fig:mode_shape_gallery} and \ref{fig:mac_violin} indicate that the proposed framework provides reliable reconstruction of spatial modal patterns at both the individual-sample and full-dataset levels.

\subsection{Global Prediction Consistency Across Modes}

\begin{figure}[t]
	\centering
	\begin{subfigure}[b]{0.95\linewidth}
		\centering
		\includegraphics[width=\linewidth]{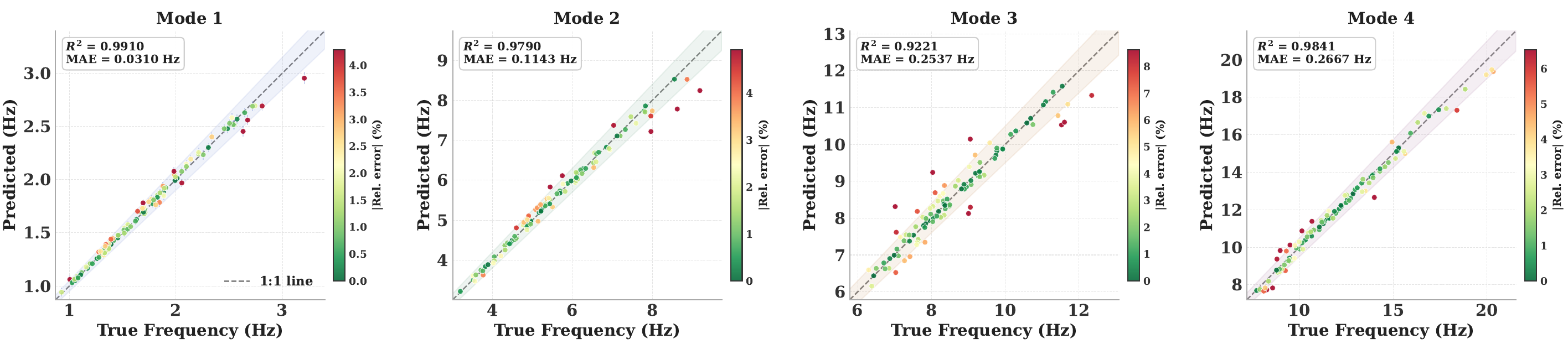}
		\caption{Natural frequency: true vs predicted}
		\label{fig:freq_scatter}
	\end{subfigure}
	\hfill
	\begin{subfigure}[b]{0.95\linewidth}
		\centering
		\includegraphics[width=\linewidth]{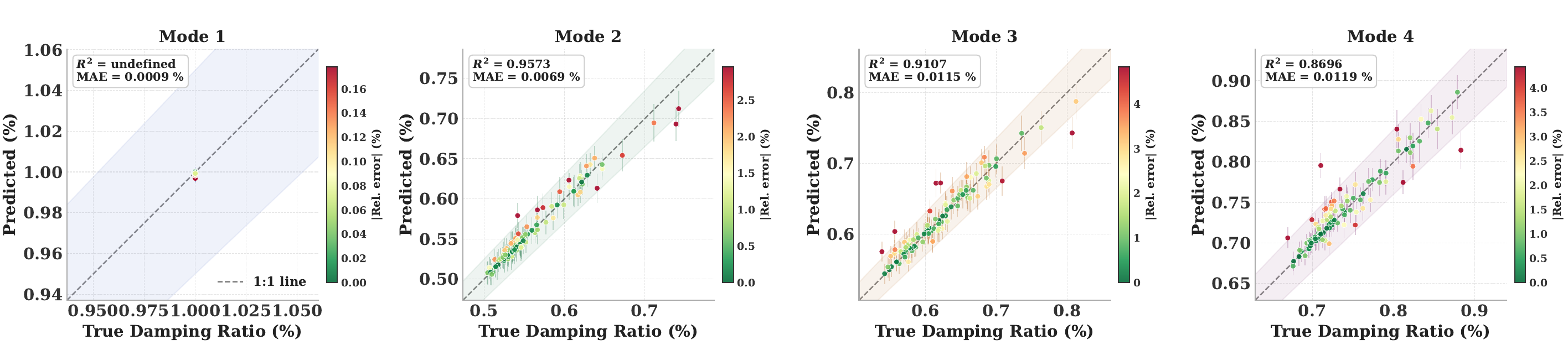}
		\caption{Damping ratio: true vs predicted}
		\label{fig:zeta_scatter}
	\end{subfigure}
	\caption{Predicted-versus-true scatter plots for the global modal parameters over the test set. Panel (a) shows the natural frequency predictions and panel (b) shows the damping ratio predictions. In both cases, close alignment with the 1:1 line and concentration within the $\pm 10\%$ bounds indicate strong predictive accuracy and limited bias.}
	\label{fig:scatter_comparison}
\end{figure}

Additional insight into the global modal-parameter predictions is provided through the predicted-versus-true scatter plots in Fig.~\ref{fig:scatter_comparison}. For both natural frequencies and damping ratios, the predictions align closely with the 1:1 line, indicating low bias and strong correspondence with the ground truth. A large proportion of the predictions lie within the $\pm 10\%$ error bounds, demonstrating consistent predictive accuracy across all modal parameters. The agreement is especially strong for the lower modes, while slightly larger dispersion is visible for the higher modes, in accordance with the error distributions discussed previously.

Table~\ref{tab:modal_stats} summarises the statistical performance across all modes. The mean MAC values remain above 0.96 for all modes, with the strongest agreement observed for Modes 1 and 2. Frequency errors remain small overall, although the spread increases for the higher modes, particularly for Mode 3. Damping errors exhibit larger variability than frequency errors, especially for Modes 3 and 4, which is consistent with the greater difficulty of damping estimation. Overall, the results indicate that the proposed model provides accurate and stable predictions across all modal parameters, while exhibiting the expected reduction in performance for higher-order modes.

\begin{table}[H]
	\centering
	\caption{Statistical performance across modes.}
	\label{tab:modal_stats}
	\begin{tabular}{c|ccc|ccc|ccc}
		\hline
		\multirow{2}{*}{\textbf{Mode number}}
		& \multicolumn{3}{c|}{\textbf{MAC}}
		& \multicolumn{3}{c|}{\textbf{Damping Error (\%)}}
		& \multicolumn{3}{c}{\textbf{Frequency Error (\%)}} \\
		\cline{2-10}
		& Mean & Std & Max
		& Mean & Std & Max
		& Mean & Std & Max \\
		\hline
		Mode 1 & 0.9945 & 0.0040 & 0.9776 & -0.09 & 0.058 & 0.314 & 0.386 & 2.204 & 8.129 \\
		Mode 2 & 0.9888 & 0.0111 & 0.9278 & 0.241 & 1.652 & 6.939 & -0.137 & 2.711 & 10.122 \\
		Mode 3 & 0.9689 & 0.0618 & 0.4680 & 0.465 & 2.552 & 9.286 & 0.240 & 4.469 & 18.861 \\
		Mode 4 & 0.9749 & 0.0494 & 0.6602 & 0.550 & 2.303 & 12.127 & -0.450 & 3.024 & 9.819 \\
		\hline
	\end{tabular}
\end{table}

\subsection{Uncertainty Decomposition}

\begin{figure}[H]
	\centering
	\includegraphics[width=0.95\linewidth]{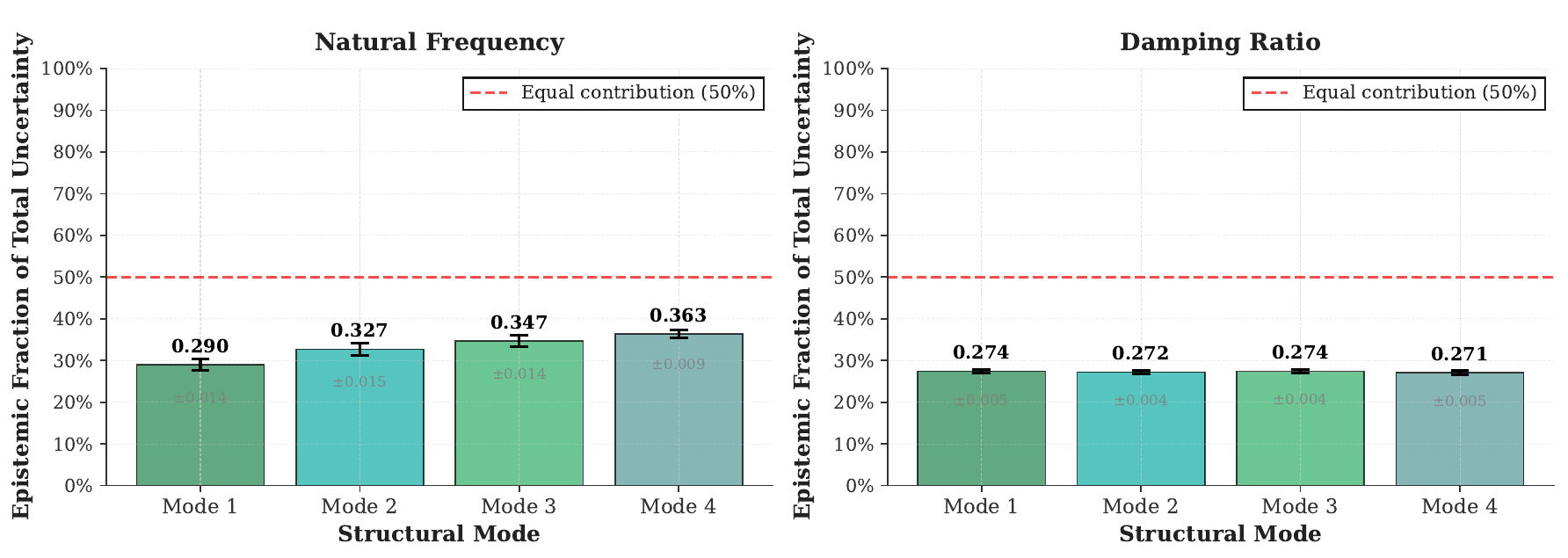}
	\caption{Quantitative contribution of epistemic and aleatoric uncertainty across the first four modes for frequency and damping ratio estimation.}
	\label{fig:uncertainty_decomposition}
\end{figure}

The predictive uncertainty is further analysed through its decomposition into epistemic and aleatoric components. Fig.~\ref{fig:uncertainty_decomposition} shows the relative contributions of these two sources of uncertainty across different modes and test conditions. In most cases, aleatoric uncertainty constitutes the dominant component. For frequency estimation, the epistemic fraction increases from approximately 0.29 for Mode 1 to 0.36 for Mode 4, which indicates that aleatoric uncertainty still accounts for nearly 64--71\% of the total variance. For damping estimation, the epistemic contribution remains nearly constant at around 0.27 across all modes, implying that approximately 73\% of the uncertainty is aleatoric. This predominance of aleatoric uncertainty suggests that measurement noise remains the primary source of variability.

A more detailed examination shows that, for frequency estimation, the epistemic fraction increases progressively from 0.290 in Mode 1 to 0.327, 0.347, and 0.363 for Modes 2--4, respectively. In contrast, for damping estimation, the epistemic contribution remains nearly constant at approximately 0.271--0.274 across all modes. This indicates that epistemic uncertainty becomes more pronounced for higher modes, particularly in frequency prediction. Such behaviour is consistent with the greater difficulty of identifying higher-order dynamics, where modal responses are weaker and more sensitive to noise. A similar trend would be expected under reduced sensor availability, where limited spatial information would further increase epistemic uncertainty through incomplete observability of the structural response.

This decomposition provides additional insight into model behaviour. Cases with higher prediction error are generally associated with increased total uncertainty, with a noticeable contribution from the epistemic component. This suggests that the model reflects its confidence appropriately in regions where the data are less informative. The distinct behaviour of aleatoric and epistemic components across modes further indicates that the model is not merely producing aggregate uncertainty, but is effectively separating noise-driven variability from structural and data-dependent uncertainty.

\subsection{Predictive Distribution and Uncertainty Behaviour}

\begin{figure}[H]
	\centering
	\begin{subfigure}[b]{0.95\linewidth}
		\centering
		\includegraphics[width=\linewidth]{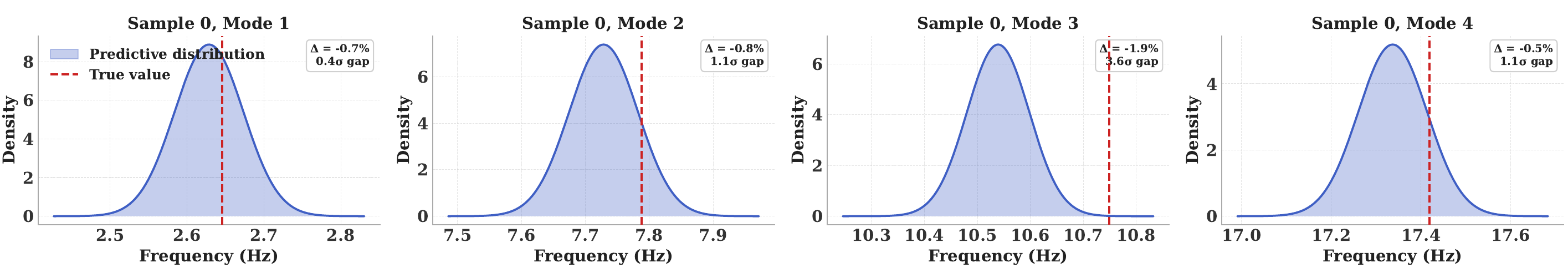}
		\caption{Natural frequency: true vs predicted}
		\label{fig:uq_frequency_grid}
	\end{subfigure}
	\hfill
	\begin{subfigure}[b]{0.95\linewidth}
		\centering
		\includegraphics[width=\linewidth]{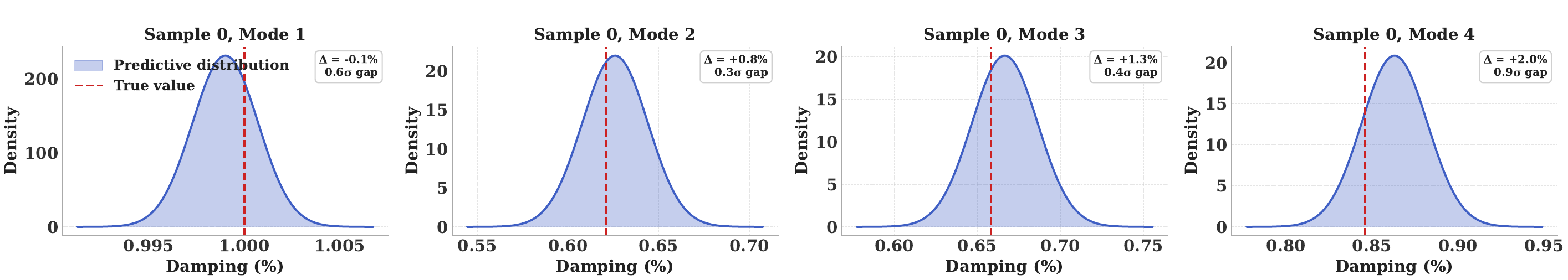}
		\caption{Damping ratio: true vs predicted}
		\label{fig:uq_damping_grid}
	\end{subfigure}
	\caption{Predictive distributions of modal parameters for representative test samples. The Gaussian curves represent the predicted probability density, with the dotted blue line indicating the predicted mean and the dashed red line denoting the ground truth. \textbf{Top row}: natural frequencies (Hz); \textbf{bottom row}: damping ratios (\%).}
	\label{fig:uncertainty_modal}
\end{figure}

The proposed framework provides predictive distributions for modal parameters rather than deterministic point estimates. Representative results are shown in Fig.~\ref{fig:uncertainty_modal}, where each prediction is expressed as a Gaussian distribution characterised through its mean and variance.

The predicted mean values are generally well aligned with the ground truth across all modes. For cases with low prediction error, the corresponding distributions are narrow, indicating high confidence. In contrast, wider distributions are observed in more challenging cases, particularly for higher modes and for damping estimation. This behaviour suggests that the model adjusts its uncertainty in response to prediction difficulty. In most cases, the ground truth lies within high-probability regions of the predicted distributions, which indicates consistency between the predicted mean and the associated uncertainty.

The results also show that the spread of the predictive distributions increases for higher-order modes, which is consistent with their reduced observability and increased sensitivity to noise. Similarly, damping predictions exhibit larger variance than frequency predictions, reflecting their greater estimation difficulty. Overall, the proposed model provides coherent probabilistic predictions in which both the mean and variance contribute to a more informative representation of modal parameters.

\subsection{Calibration and Reliability Assessment}

\begin{figure}[t]
	\centering
	\includegraphics[width=1\linewidth]{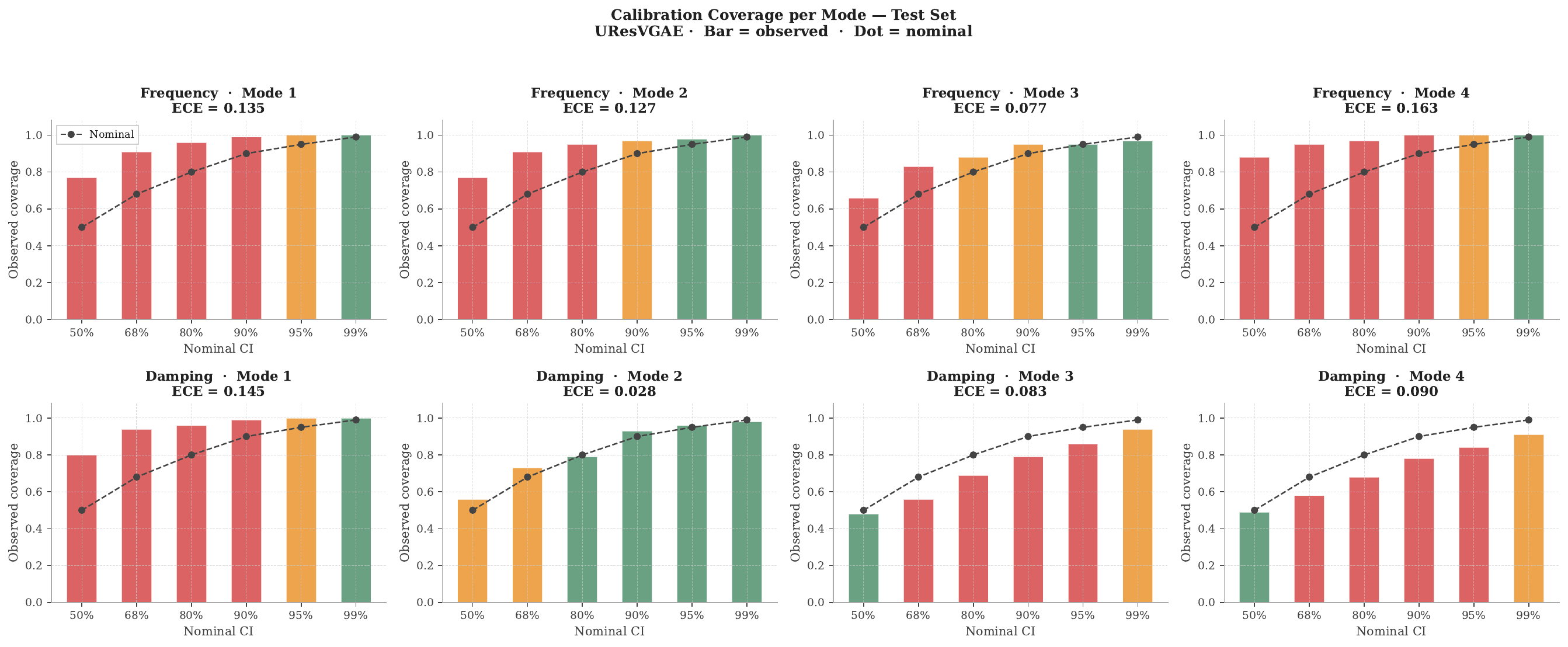}
	\caption{Calibration performance across confidence levels for frequency (\textbf{top row}) and damping ratio (\textbf{bottom row}) predictions. Bars represent the observed empirical coverage for each nominal confidence interval, while the dashed line with markers indicates the ideal (nominal) coverage. Good calibration is reflected by close agreement between observed and nominal values. The proposed model shows generally consistent calibration across modes, with minor deviations observed for higher confidence levels and damping predictions. The ECE for each mode is reported, indicating overall reliable uncertainty estimation.}
	\label{fig:coverage_bars}
\end{figure}

The quality of the predicted uncertainty is evaluated through
calibration analysis, which assesses the consistency between predicted
confidence intervals and empirical coverage. A well-calibrated model
is expected to produce confidence intervals that match the true
frequency of occurrence
\cite{guo2017calibration,pavlovic2025understanding}.
Fig.~\ref{fig:coverage_bars} presents the reliability diagrams for
both frequency and damping predictions. The observed coverage closely
follows the ideal diagonal, indicating good calibration. The ECE for
frequency ranges from 0.077 to 0.163 across Modes 1--4, while for
damping it varies between 0.028 and 0.145. Notably, the calibration
error does not follow a monotonic trend with mode order: Mode 3 shows
the lowest ECE (0.077) for frequency, whereas Mode 4 exhibits a higher
deviation (0.163); for damping, Mode 2 achieves the lowest ECE
(0.028), while Mode 1 is higher (0.145). This behaviour indicates that
calibration is not governed solely by modal complexity, but also by
how well the predicted uncertainty aligns with the empirical error
distribution.

\begin{figure}[H]
	\centering
	\includegraphics[width=0.9\linewidth]{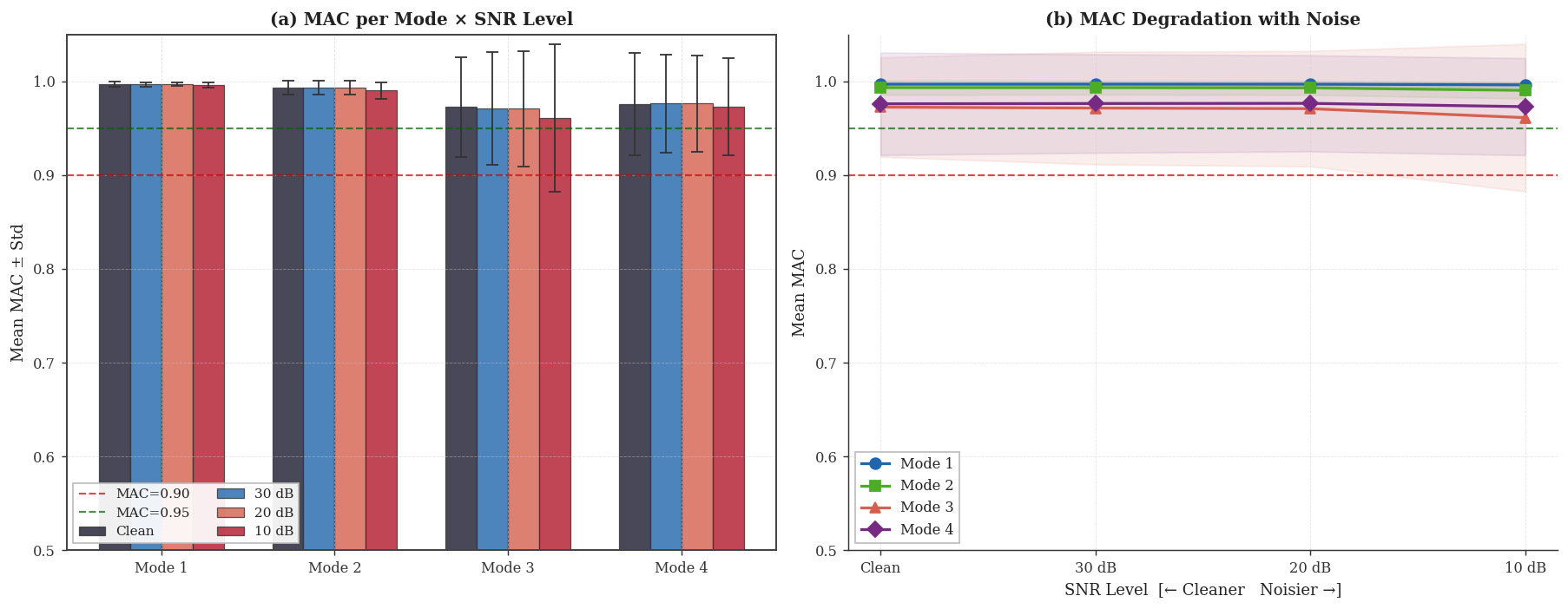}
	\caption{Distribution of MAC values for Modes 1--4 under varying noise conditions (Clean, 30 dB, 20 dB, and 10 dB SNR). The boxplots illustrate the stability of mode-shape identification, while the shaded density bands on the right of each subplot provide a detailed view of the predictive distribution and the frequency of high-fidelity (MAC $> 0.9$) identifications.}
	\label{fig:noise_mac}
\end{figure}

In particular, lower ECE can be achieved even in more challenging modes
when the uncertainty estimates are well matched to prediction
variability, whereas slight misalignment can lead to higher ECE even
for easier modes. Minor deviations from the diagonal are observed at
higher confidence levels, particularly for damping estimation, which
can be attributed to its greater sensitivity to noise. Nevertheless,
the overall trend remains consistent, and no systematic overconfidence
or underconfidence is evident.

\subsection{Comparison with Baseline GNN Framework}

To further evaluate the robustness of the proposed framework under realistic measurement conditions, a comparative analysis is performed against a baseline GNN model. The assessment focuses on the influence of measurement noise and sparse sensing on modal parameter estimation, thereby enabling a systematic evaluation of the relative stability and robustness of the proposed architecture, particularly for higher modes and more challenging identification scenarios.

\subsubsection{Performance under Measurement Noise}

\begin{figure}[H]
	\centering
	\includegraphics[width=0.9\linewidth]{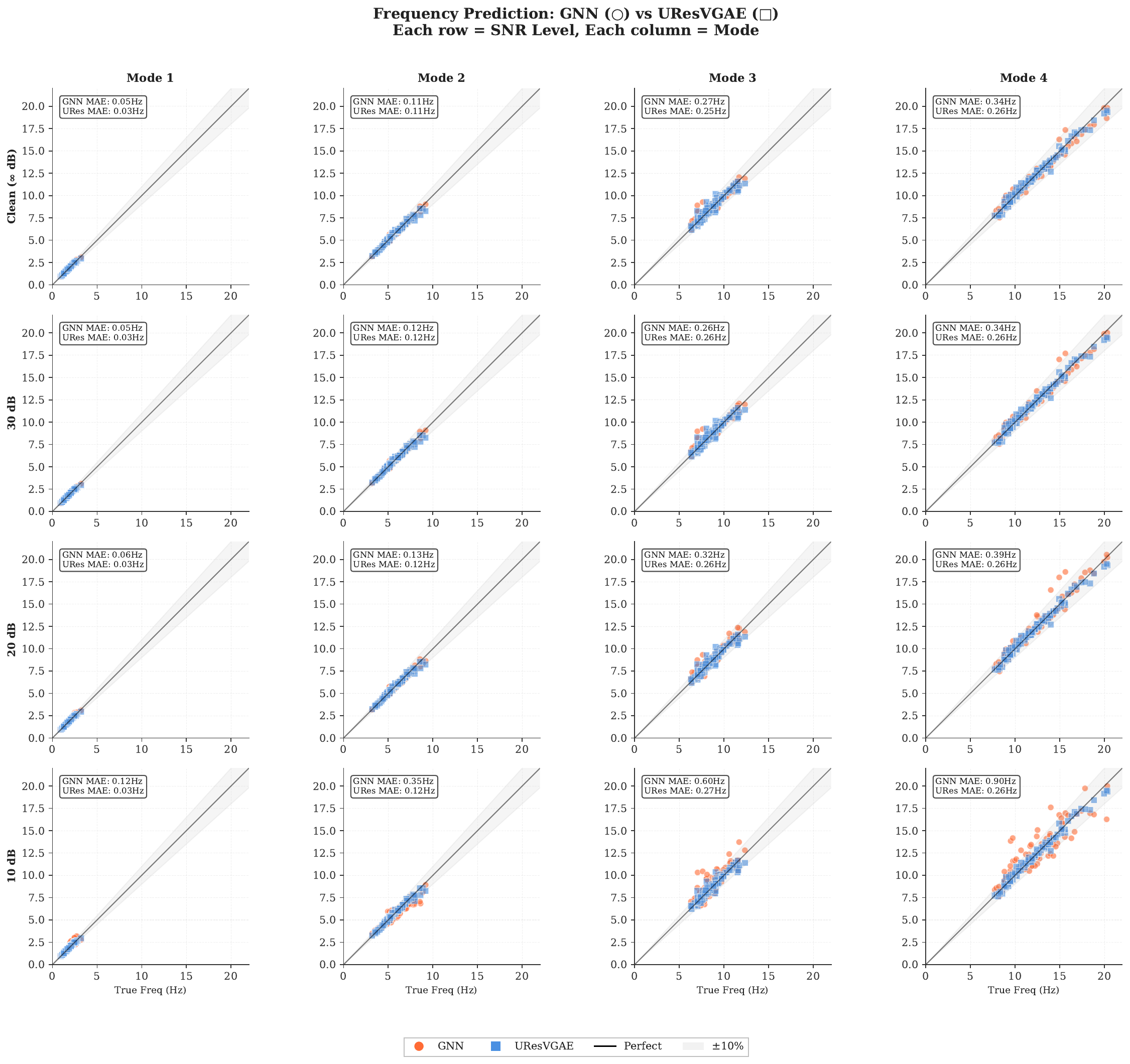}
	\caption{Comparison of natural frequency prediction performance between the baseline GNN and the proposed UResVGAE under varying noise conditions (Clean, 30 dB, 20 dB, and 10 dB SNR) for Modes 1–4. Each subplot presents predicted versus true frequencies, with corresponding mean absolute error (MAE) values indicated. Under clean conditions, both models exhibit close agreement with the ground truth. However, as noise levels increase, the baseline GNN shows increased dispersion and deviation from the ideal 1:1 relationship, particularly for higher modes. In contrast, UResVGAE maintains tighter clustering and lower error, demonstrating improved robustness to measurement noise and more stable performance across modal orders.}
	\label{fig:model_comparison_frequency}
\end{figure}

Table~\ref{tab:avg_comparison} summarises the average performance across all modes for UResVGAE and the baseline GNN under different noise levels. Across all SNR conditions, UResVGAE maintains higher mean MAC values and lower frequency and damping errors than the baseline model. The performance gap becomes more pronounced as the noise level increases, indicating improved robustness of the proposed framework under degraded measurement conditions. This trend is also reflected in Fig.~\ref{fig:noise_mac}, which shows the variation in MAC values for Modes 1--4 under different noise conditions. Mode-shape identification remains stable across the considered SNR levels, with only modest degradation as the noise increases. The decline is more visible for the higher modes, but the overall MAC values remain high, indicating that the proposed framework preserves strong mode-shape fidelity even under noisy measurements.

\begin{table}[H]
	\centering
	\caption{Average performance across all modes: UResVGAE vs GNN.}
	\label{tab:avg_comparison}
	\begin{tabular}{l c c c c c c}
		\toprule
		\multirow{2}{*}{\textbf{SNR (dB)}} & \multicolumn{2}{c}{\textbf{MAC (Mean)}} & \multicolumn{2}{c}{\textbf{Freq Error MAE (\%)}} & \multicolumn{2}{c}{\textbf{Damp Error MAE (\%)}} \\
		\cmidrule(lr){2-3} \cmidrule(lr){4-5} \cmidrule(lr){6-7}
		& UResVGAE & GNN & UResVGAE & GNN & UResVGAE & GNN \\
		\midrule
		Clean & \textbf{0.9819} & 0.9712 & \textbf{0.301} & 2.69 & \textbf{0.339} & 1.74 \\
		30 dB & \textbf{0.9817} & 0.9692 & \textbf{0.302} & 2.63 & \textbf{0.340} & 1.71 \\
		20 dB & \textbf{0.9813} & 0.9575 & \textbf{0.313} & 3.37 & \textbf{0.341} & 1.96 \\
		10 dB & \textbf{0.9765} & 0.8985 & \textbf{0.32} & 6.27 & \textbf{0.326} & 3.53 \\
		\bottomrule
	\end{tabular}
\end{table}

\begin{figure}[H]
	\centering
	\includegraphics[width=0.9\linewidth]{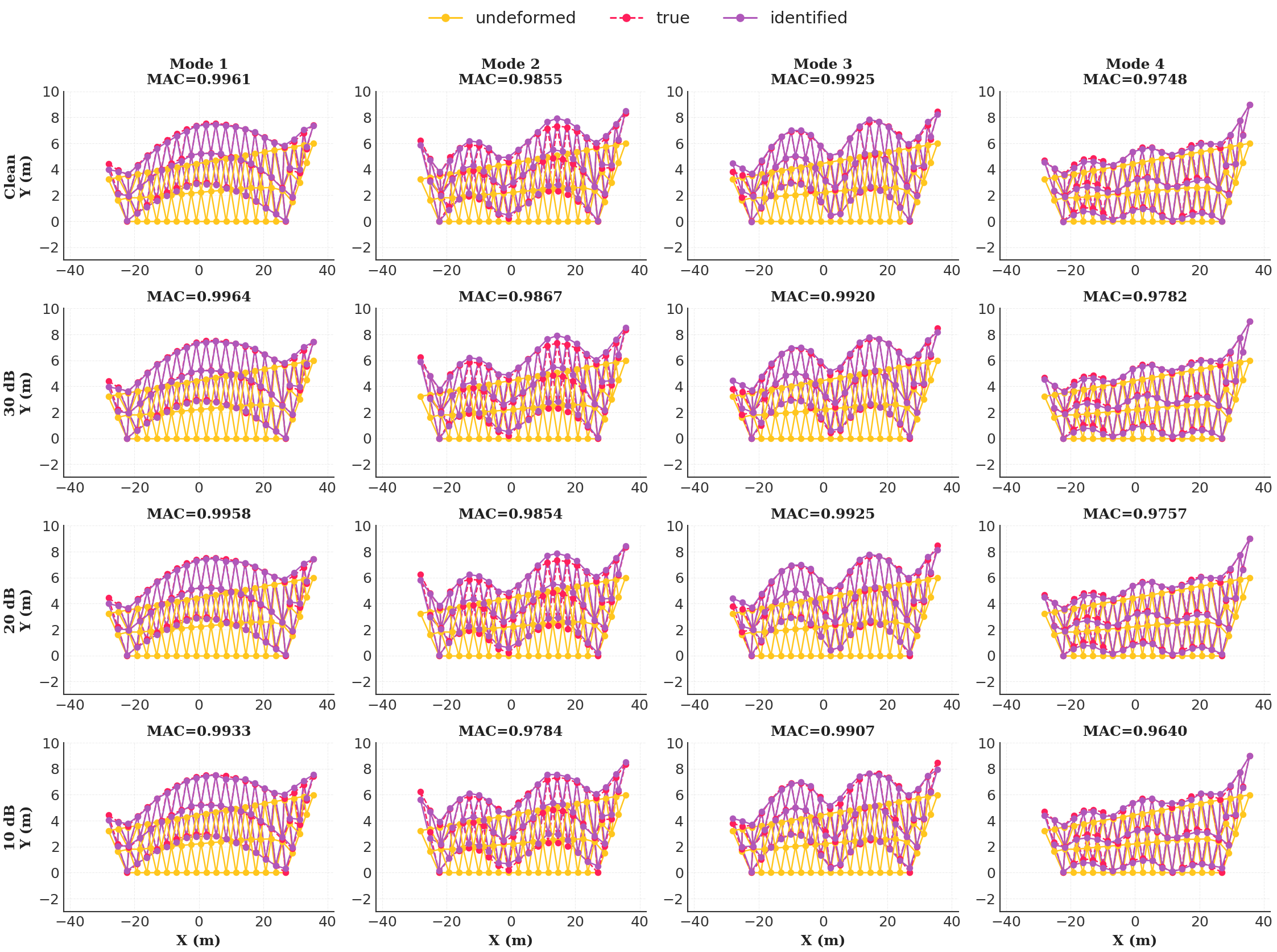}
	\caption{Comparison of predicted mode shapes obtained using the baseline GNN and the proposed UResVGAE under varying noise conditions (Clean, 30 dB, 20 dB, and 10 dB SNR) for Modes 1–4. The figure presents a total of 16 subplots, illustrating the spatial deformation patterns for each mode across different noise levels. Under clean conditions, both models produce mode shapes that closely resemble the reference patterns. However, as the noise level increases, the baseline GNN exhibits noticeable distortions and loss of spatial consistency, particularly for higher modes. In contrast, UResVGAE maintains more coherent and physically consistent mode shapes across all SNR levels, demonstrating improved robustness to noise and better preservation of structural dynamics.}
	\label{fig:modeshape_comp}
\end{figure}

Fig.~\ref{fig:model_comparison_frequency} compares the frequency-prediction performance of the baseline GNN and the proposed UResVGAE across different SNR levels. While both models perform well under clean conditions, the proposed method maintains lower error and tighter clustering around the diagonal as noise increases. The performance gap becomes more pronounced for higher modes and at lower SNR levels, particularly at 10 dB, where the baseline GNN exhibits increased dispersion and larger deviations from the ground truth. These results indicate that the proposed model is more robust to noise and better captures the underlying structural dynamics under degraded measurement conditions.

The corresponding comparison of the predicted mode shapes for UResVGAE and the baseline GNN under different SNR levels is presented in Fig.~\ref{fig:modeshape_comp}. The visual agreement between the predicted and undeformed reference shapes remains consistently strong for UResVGAE across all noise levels. In contrast, the baseline GNN shows visibly greater deviations in some of the higher modes as noise increases. This further supports the conclusion that the proposed framework offers improved robustness in both global modal-parameter estimation and spatial mode-shape reconstruction.

\begin{figure}[H]
	\centering
	\includegraphics[width=1\linewidth]{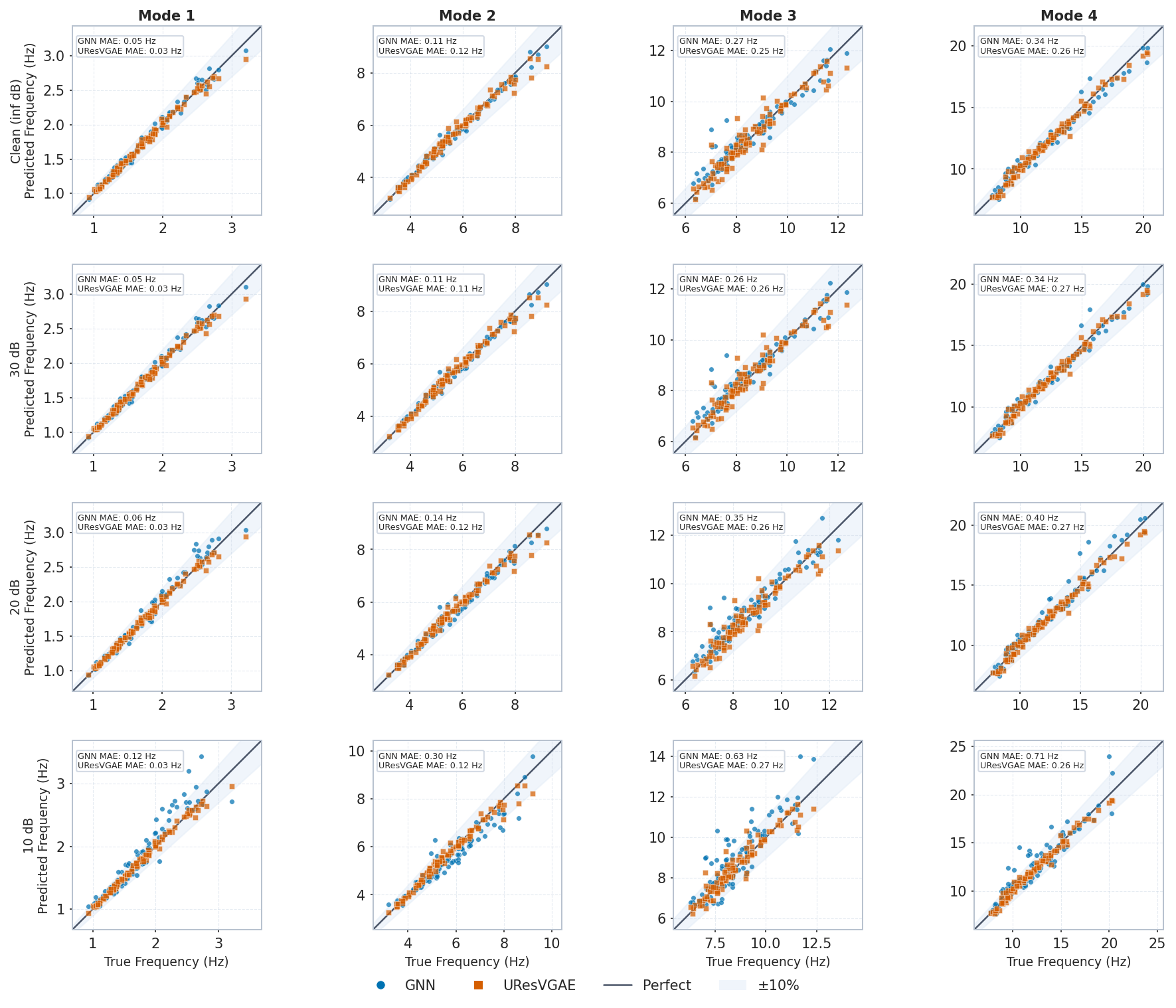}
	\caption{Per-mode frequency prediction scatter plots for GNN and UResVGAE across all SNR levels. Each row corresponds to an SNR level and each column to a vibration mode; points closer to the diagonal indicate more accurate predictions, with UResVGAE generally exhibiting smaller deviations, particularly in higher modes and noisier cases.}
	\label{fig:scatter1}
\end{figure}

\subsubsection{Performance under Sensor Sparsity Investigation}

To examine the effect of reduced sensor availability on frequency estimation, Fig.~\ref{fig:scatter1} presents per-mode prediction scatter plots for both GNN and UResVGAE across all sensing conditions. The plots compare predicted values against the corresponding ground truth, thereby providing insight into accuracy and dispersion. This representation facilitates a mode-wise evaluation of model behaviour under varying levels of sensor sparsity. The predictions for both models generally align well with the ideal diagonal, particularly for the lower modes and higher sensor fractions. The panel-wise arrangement makes it possible to observe how prediction quality changes simultaneously with mode number and sensor availability. For the higher modes, especially under stronger sensing reduction, the GNN predictions show a wider spread around the reference line. The UResVGAE predictions remain more concentrated, indicating better consistency in frequency estimation under sparse sensing.

\begin{figure}[H]
	\centering
	\includegraphics[width=1\linewidth]{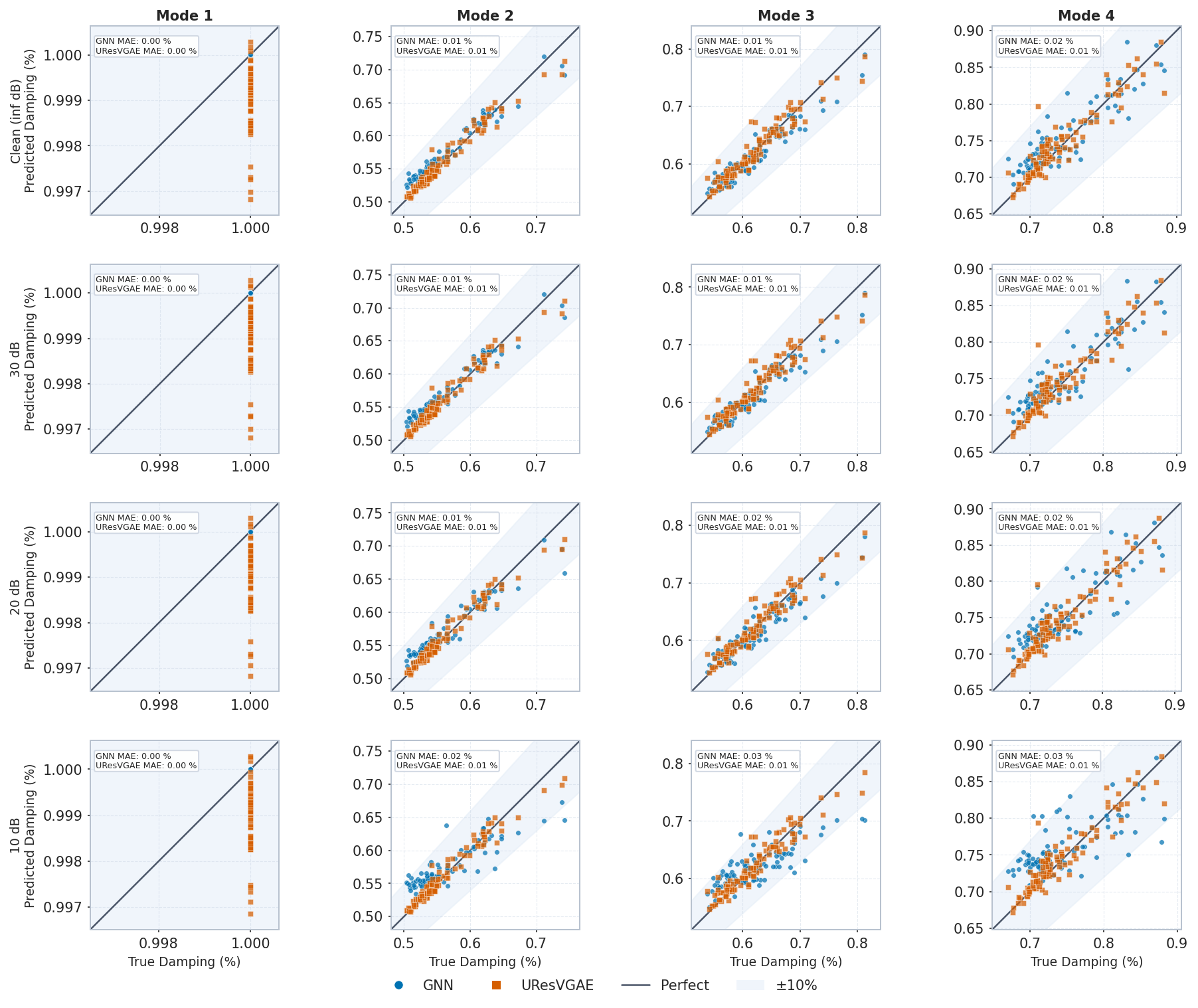}
	\caption{Per-mode damping prediction scatter plots for GNN and UResVGAE across all SNR levels. Each row corresponds to an SNR level and each column to a vibration mode; both models track the reference trend, while UResVGAE shows tighter clustering around the ideal line in the more challenging modes and noise conditions.}
	\label{fig:scatter2}
\end{figure}

Fig.~\ref{fig:scatter2} presents the corresponding per-mode damping prediction results. Compared with frequency, damping estimation shows greater dispersion, which is expected since damping is typically more sensitive to modelling and measurement uncertainty. This effect becomes more pronounced in the higher modes and at lower sensor fractions, where the GNN predictions exhibit larger deviations from the ideal trend. In most panels, UResVGAE produces a tighter clustering around the diagonal, which points to a more reliable damping-identification capability under sparse sensing.

\begin{table}[t]
	\centering
	\caption{Comparison of GNN and UResVGAE in the low-sensor FPA setting using values extracted from the saved \texttt{.mat} result files. For MAC, higher values are better; for frequency and damping MAE, lower values are better.}
	\label{tab:low_sensor_fpa}
	\begin{tabular}{c c c c c c c}
		\toprule
		\multirow{2}{*}{\textbf{Sensors (\%)}} & \multicolumn{2}{c}{\textbf{Mean MAC}} & \multicolumn{2}{c}{\textbf{Frequency MAE (\%)}} & \multicolumn{2}{c}{\textbf{Damping MAE (\%)}} \\
		\cmidrule(lr){2-3} \cmidrule(lr){4-5} \cmidrule(lr){6-7}
		& GNN & UResVGAE & GNN & UResVGAE & GNN & UResVGAE \\
		\midrule
		5  & 0.7181 & \textbf{0.7354} & 9.2501 & \textbf{7.1616} & 4.6772 & \textbf{3.2964} \\
		10 & \textbf{0.8006} & 0.7943 & 6.9256 & \textbf{4.7214} & 4.3886 & \textbf{2.7644} \\
		20 & \textbf{0.8873} & 0.8715 & 4.6814 & \textbf{3.1521} & 3.5337 & \textbf{2.0130} \\
		30 & \textbf{0.9331} & 0.9172 & 4.0073 & \textbf{2.7009} & 3.0893 & \textbf{1.6586} \\
		50 & \textbf{0.9579} & 0.9525 & 3.2704 & \textbf{2.4221} & 2.2093 & \textbf{1.3837} \\
		80 & 0.9695 & \textbf{0.9736} & 2.7808 & \textbf{2.2840} & 1.8897 & \textbf{1.1807} \\
		95 & 0.9712 & \textbf{0.9801} & 2.6985 & \textbf{2.2172} & 1.7543 & \textbf{1.1766} \\
		\midrule
		Average & \textbf{0.8911} & 0.8892 & 4.8020 & \textbf{3.5228} & 3.0774 & \textbf{1.9248} \\
		\bottomrule
	\end{tabular}
\end{table}

Table~\ref{tab:low_sensor_fpa} summarises the low-sensor FPA performance of GNN and UResVGAE across different sensor-availability levels. Although the mean MAC values of the two models are broadly comparable, UResVGAE consistently achieves lower frequency and damping errors at every sensor fraction considered. The reduction in frequency MAE is particularly notable in the sparse-sensing regime, while the improvement in damping MAE remains strong across the full range of sensor availability. These results indicate that UResVGAE provides more accurate modal-parameter estimation even when the overall modal correlation is similar to that of the baseline GNN.

Robustness to sparse sensing is assessed by progressively reducing the fraction of observed nodes. Even when the available sensors are reduced to 20--30\% of the full configuration, UResVGAE maintains high modal correlation, with MAC values remaining above 0.90 across most modes. At the same time, frequency and damping errors show only a moderate increase, remaining within approximately 3--5\% and 2--4\%, respectively. This demonstrates that the proposed framework retains reliable performance even under significantly limited sensor availability.

\section{Conclusion}
\label{sec:conclusion}

This work presented a physics-aware variational graph-based framework,
UResVGAE, for probabilistic modal identification from power spectral
density (PSD) data under noisy and incomplete measurement conditions.
The proposed architecture combines residual graph neural networks with
a variational latent representation to jointly estimate natural
frequencies, damping ratios, and spatially distributed mode shapes
within a unified formulation. A hybrid uncertainty-quantification
strategy based on evidential regression \cite{amini2020deep} and
Bayesian approximation \cite{maddox2019simple,gal2015dropout} enables
the model to produce calibrated predictive distributions.

The uncertainty analysis provides further insight into model behaviour. Aleatoric uncertainty dominates across most conditions, accounting for approximately 65--75\% of the total predictive variance, particularly in frequency estimation. In contrast, the epistemic component increases with modal order, from approximately 0.29 to 0.36 for frequency, reflecting reduced observability and greater modelling difficulty for higher modes. Calibration results further show that the predicted confidence intervals remain well aligned with empirical coverage, with ECE values lying within 0.028--0.163 across all modes. The absence of a monotonic trend in calibration error indicates that reliability is governed not only by modal complexity, but also by the extent to which the predicted uncertainty matches the empirical error distribution.

The proposed framework also demonstrates improved robustness under practical constraints. Under increasing noise levels, UResVGAE maintains higher MAC values and lower prediction errors than the baseline GNN, with the performance gap becoming more pronounced at lower SNR levels. Under sparse sensing conditions, the model preserves strong modal correlation and maintains frequency and damping errors within a limited range even when only 20--30\% of nodes are observed. These results indicate that the learned representations capture global structural behaviour effectively from partial and noisy observations. Overall, the proposed UResVGAE framework provides a unified and physically consistent approach for modal identification with calibrated uncertainty estimation. Through the joint modelling of global modal parameters and node-level mode shapes within a probabilistic graph-based architecture, the method addresses important limitations of existing approaches related to noise sensitivity, data sparsity, and limited uncertainty awareness, while also establishing a clear foundation for future validation on experimental and real-world structural datasets.

\bibliographystyle{unsrt}  
\bibliography{references}

\end{document}